\documentclass[twocolumn]{aastex701}
\usepackage{amsmath}
\usepackage{comment}

\graphicspath{{./}{figures/}}

\newcommand{\logmass}{\ensuremath{\log_{10}(M_*/M_{\odot})}}

\defcitealias{Thornton2024}{T24}

\begin{document}
\title{
Dwarf galaxy halo masses from spectroscopic and photometric lensing in DESI and DES
}
\author[0000-0003-0660-9776]{Helena Treiber}
\affiliation{Department of Astrophysical Sciences, Princeton University, 4 Ivy Lane, Princeton, NJ 08544, USA}
\email{lena.treiber@princeton.edu}

\author[0000-0002-6445-0559]{Alexandra Amon}
\affiliation{Department of Astrophysical Sciences, Princeton University, 4 Ivy Lane, Princeton, NJ 08544, USA}
\email{alexandra.amon@princeton.edu}

\author[0000-0003-2229-011X]{Risa H. Wechsler}
\affiliation{Kavli Institute for Particle Astrophysics and Cosmology and Department of Physics, Stanford University, Stanford, CA 94305, USA}
\affiliation{SLAC National Accelerator Laboratory, Menlo Park, CA 94025, USA}
\email{}

\author[0000-0002-7113-0262]{Viraj Manwadkar}
\affiliation{Kavli Institute for Particle Astrophysics and Cosmology and Department of Physics, Stanford University, Stanford, CA 94305, USA}
\affiliation{SLAC National Accelerator Laboratory, Menlo Park, CA 94025, USA}
\email{}

\author[0000-0001-6145-5859]{Justin Myles}
\affiliation{Department of Astrophysical Sciences, Princeton University, 4 Ivy Lane, Princeton, NJ 08544, USA}
\email{}

\author[0009-0004-9275-3559]{Joseph Thornton}
\affiliation{DAMTP, Centre for Mathematical Sciences, Wilberforce Road, Cambridge CB3 0WA, UK}
\email{}

\author[0000-0002-1200-0820]{Yao-Yuan Mao}
\affiliation{Department of Physics and Astronomy, University of Utah, Salt Lake City, UT 84112, USA}
\email{}

\author[0000-0003-1197-0902]{ChangHoon Hahn}
\affiliation{Steward Observatory, University of Arizona, 933 N. Cherry Avenue, Tucson, AZ 85721, USA}
\email{}

\author[]{Andrew Hearin}
\affiliation{Argonne National Laboratory, High-Energy Physics Division, 9700 S. Cass Avenue, Argonne, IL 60439, USA}
\email{}

\author[0000-0002-7273-4076]{Sven Heydenreich}
\affiliation{Department of Astronomy and Astrophysics, UCO, University of California, 1156 High Street, Santa Cruz, CA 95064, USA}
\email{}

\author[0000-0003-4357-3450]{Amélie Saintonge}
\affiliation{Department of Physics \& Astronomy, University College London, Gower Street, London, WC1E 6BT, UK}
\email{}

\author[0000-0003-4755-6404]{Manasvee Saraf}
\affiliation{Department of Physics \& Astronomy, University College London, Gower Street, London, WC1E 6BT, UK}
\email{}

\author[]{Jessica Nicole Aguilar}
\affiliation{Lawrence Berkeley National Laboratory, 1 Cyclotron Road, Berkeley, CA 94720, USA}
\email{}

\author[0000-0001-6098-7247]{Steven Ahlen}
\affiliation{Department of Physics, Boston University, 590 Commonwealth Avenue, Boston, MA 02215 USA}
\email{}

\author[0000-0003-2923-1585]{Abhijeet Anand}
\affiliation{Lawrence Berkeley National Laboratory, 1 Cyclotron Road, Berkeley, CA 94720, USA}
\email{}

\author[0000-0001-9712-0006]{Davide Bianchi}
\affiliation{Dipartimento di Fisica ``Aldo Pontremoli'', Universit\`a degli Studi di Milano, Via Celoria 16, I-20133 Milano, Italy}
\affiliation{INAF-Osservatorio Astronomico di Brera, Via Brera 28, 20122 Milano, Italy}
\email{}

\author[]{David Brooks}
\affiliation{Department of Physics \& Astronomy, University College London, Gower Street, London, WC1E 6BT, UK}
\email{}

\author[0000-0001-7316-4573]{Francisco Javier Castander}
\affiliation{Institut d'Estudis Espacials de Catalunya (IEEC), c/ Esteve Terradas 1, Edifici RDIT, Campus PMT-UPC, 08860 Castelldefels, Spain}
\affiliation{Institute of Space Sciences, ICE-CSIC, Campus UAB, Carrer de Can Magrans s/n, 08913 Bellaterra, Barcelona, Spain}
\email{}

\author[]{Todd Claybaugh}
\affiliation{Lawrence Berkeley National Laboratory, 1 Cyclotron Road, Berkeley, CA 94720, USA}
\email{}

\author[0000-0001-8274-158X]{Andrew P. Cooper}
\affiliation{Institute of Astronomy and Department of Physics, National Tsing Hua University, 101 Kuang-Fu Rd. Sec. 2, Hsinchu 30013, Taiwan}
\email{}

\author[0000-0002-2169-0595]{Andrei Cuceu}
\affiliation{Lawrence Berkeley National Laboratory, 1 Cyclotron Road, Berkeley, CA 94720, USA}
\email{}

\author[0000-0002-1769-1640]{Axel de la Macorra}
\affiliation{Instituto de F\'{\i}sica, Universidad Nacional Aut\'{o}noma de M\'{e}xico, Circuito de la Investigaci\'{o}n Cient\'{\i}fica, Ciudad Universitaria, Cd. de M\'{e}xico  C.~P.~04510, M\'{e}xico}
\email{}

\author[0000-0002-5665-7912]{Biprateep Dey}
\affiliation{Department of Astronomy \& Astrophysics, University of Toronto, Toronto, ON M5S 3H4, Canada}
\affiliation{Department of Physics \& Astronomy and Pittsburgh Particle Physics, Astrophysics, and Cosmology Center (PITT PACC), University of Pittsburgh, 3941 O'Hara Street, Pittsburgh, PA 15260, USA}
\email{}

\author[0000-0002-2890-3725]{Jaime E. Forero-Romero}
\affiliation{Departamento de F\'isica, Universidad de los Andes, Cra. 1 No. 18A-10, Edificio Ip, CP 111711, Bogot\'a, Colombia}
\affiliation{Observatorio Astron\'omico, Universidad de los Andes, Cra. 1 No. 18A-10, Edificio H, CP 111711 Bogot\'a, Colombia}
\email{}

\author[0000-0001-9632-0815]{Enrique Gazta\~{n}aga}
\affiliation{Institut d'Estudis Espacials de Catalunya (IEEC), c/ Esteve Terradas 1, Edifici RDIT, Campus PMT-UPC, 08860 Castelldefels, Spain}
\affiliation{Institute of Cosmology and Gravitation, University of Portsmouth, Dennis Sciama Building, Portsmouth, PO1 3FX, UK}
\affiliation{Institute of Space Sciences, ICE-CSIC, Campus UAB, Carrer de Can Magrans s/n, 08913 Bellaterra, Barcelona, Spain}
\email{}

\author[0000-0003-3142-233X]{Satya Gontcho A Gontcho}
\affiliation{Lawrence Berkeley National Laboratory, 1 Cyclotron Road, Berkeley, CA 94720, USA}
\affiliation{University of Virginia, Department of Astronomy, Charlottesville, VA 22904, USA}
\email{}

\author[]{Gaston Gutierrez}
\affiliation{Fermi National Accelerator Laboratory, PO Box 500, Batavia, IL 60510, USA}
\email{}

\author[0000-0001-6558-0112]{Dragan Huterer}
\affiliation{Department of Physics, University of Michigan, 450 Church Street, Ann Arbor, MI 48109, USA}
\affiliation{University of Michigan, 500 S. State Street, Ann Arbor, MI 48109, USA}
\email{}

\author[0000-0003-0201-5241]{Dick Joyce}
\affiliation{NSF NOIRLab, 950 N. Cherry Ave., Tucson, AZ 85719, USA}
\email{}

\author[0000-0002-0000-2394]{Stephanie Juneau}
\affiliation{NSF NOIRLab, 950 N. Cherry Ave., Tucson, AZ 85719, USA}
\email{}

\author[0000-0001-6356-7424]{Anthony Kremin}
\affiliation{Lawrence Berkeley National Laboratory, 1 Cyclotron Road, Berkeley, CA 94720, USA}
\email{}

\author[0000-0003-1838-8528]{Martin Landriau}
\affiliation{Lawrence Berkeley National Laboratory, 1 Cyclotron Road, Berkeley, CA 94720, USA}
\email{}

\author[0000-0001-7178-8868]{Laurent Le Guillou}
\affiliation{Sorbonne Universit\'{e}, CNRS/IN2P3, Laboratoire de Physique Nucl\'{e}aire et de Hautes Energies (LPNHE), FR-75005 Paris, France}
\email{}

\author[0000-0003-4962-8934]{Marc Manera}
\affiliation{Departament de F\'{i}sica, Serra H\'{u}nter, Universitat Aut\`{o}noma de Barcelona, 08193 Bellaterra (Barcelona), Spain}
\affiliation{Institut de F\'{i}sica d’Altes Energies (IFAE), The Barcelona Institute of Science and Technology, Edifici Cn, Campus UAB, 08193, Bellaterra (Barcelona), Spain}
\email{}

\author[0000-0002-1125-7384]{Aaron Meisner}
\affiliation{NSF NOIRLab, 950 N. Cherry Ave., Tucson, AZ 85719, USA}
\email{}

\author[]{Ramon Miquel}
\affiliation{Instituci\'{o} Catalana de Recerca i Estudis Avan\c{c}ats, Passeig de Llu\'{\i}s Companys, 23, 08010 Barcelona, Spain}
\affiliation{Institut de F\'{i}sica d’Altes Energies (IFAE), The Barcelona Institute of Science and Technology, Edifici Cn, Campus UAB, 08193, Bellaterra (Barcelona), Spain}
\email{}

\author[0000-0002-2733-4559]{John Moustakas}
\affiliation{Department of Physics and Astronomy, Siena College, 515 Loudon Road, Loudonville, NY 12211, USA}
\email{}

\author[0000-0001-9070-3102]{Seshadri Nadathur}
\affiliation{Institute of Cosmology and Gravitation, University of Portsmouth, Dennis Sciama Building, Portsmouth, PO1 3FX, UK}
\email{}

\author[0000-0002-0644-5727]{Will J. Percival}
\affiliation{Department of Physics and Astronomy, University of Waterloo, 200 University Ave W, Waterloo, ON N2L 3G1, Canada}
\affiliation{Perimeter Institute for Theoretical Physics, 31 Caroline St. North, Waterloo, ON N2L 2Y5, Canada}
\affiliation{Waterloo Centre for Astrophysics, University of Waterloo, 200 University Ave W, Waterloo, ON N2L 3G1, Canada}
\email{}

\author[0000-0001-7145-8674]{Francisco Prada}
\affiliation{Instituto de Astrof\'{i}sica de Andaluc\'{i}a (CSIC), Glorieta de la Astronom\'{i}a, s/n, E-18008 Granada, Spain}
\email{}

\author[0000-0001-6979-0125]{Ignasi P\'erez-R\`afols}
\affiliation{Departament de F\'isica, EEBE, Universitat Polit\`ecnica de Catalunya, c/Eduard Maristany 10, 08930 Barcelona, Spain}
\email{}

\author[]{Graziano Rossi}
\affiliation{Department of Physics and Astronomy, Sejong University, 209 Neungdong-ro, Gwangjin-gu, Seoul 05006, Republic of Korea}
\email{}

\author[0000-0002-9646-8198]{Eusebio Sanchez}
\affiliation{CIEMAT, Avenida Complutense 40, E-28040 Madrid, Spain}
\email{}

\author[]{David Schlegel}
\affiliation{Lawrence Berkeley National Laboratory, 1 Cyclotron Road, Berkeley, CA 94720, USA}
\email{}

\author[]{Michael Schubnell}
\affiliation{Department of Physics, University of Michigan, 450 Church Street, Ann Arbor, MI 48109, USA}
\affiliation{University of Michigan, 500 S. State Street, Ann Arbor, MI 48109, USA}
\email{}

\author[0000-0002-3461-0320]{Joseph Harry Silber}
\affiliation{Lawrence Berkeley National Laboratory, 1 Cyclotron Road, Berkeley, CA 94720, USA}
\email{}

\author[]{David Sprayberry}
\affiliation{NSF NOIRLab, 950 N. Cherry Ave., Tucson, AZ 85719, USA}
\email{}

\author[0000-0003-1704-0781]{Gregory Tarl\'{e}}
\affiliation{University of Michigan, 500 S. State Street, Ann Arbor, MI 48109, USA}
\email{}

\author[]{Benjamin Alan Weaver}
\affiliation{NSF NOIRLab, 950 N. Cherry Ave., Tucson, AZ 85719, USA}
\email{}

\author[0000-0001-5381-4372]{Rongpu Zhou}
\affiliation{Lawrence Berkeley National Laboratory, 1 Cyclotron Road, Berkeley, CA 94720, USA}
\email{}

\author[0000-0002-6684-3997]{Hu Zou}
\affiliation{National Astronomical Observatories, Chinese Academy of Sciences, A20 Datun Road, Chaoyang District, Beijing, 100101, P.~R.~China}
\email{}

\begin{abstract}
We present the most precise and lowest-mass weak lensing measurements of dwarf galaxies to date, enabled by spectroscopic lenses from the Dark Energy Spectroscopic Instrument (DESI) and photometric lenses from the Dark Energy Survey (DES) calibrated with DESI redshifts. Using DESI spectroscopy from the first data release, we construct clean samples of galaxies with median stellar masses \logmass~$=8.3–10.1$ and measure their weak lensing signals with sources from DES, KiDS, and SDSS, achieving detections with $S/N$ up to 14 for dwarf galaxies (\logmass$<$9.25)---opening up a new regime for lensing measurements of low-mass systems.
Leveraging DES photometry calibrated with DESI, we extend to a photometric dwarf sample of over 700,000 galaxies, enabling robust lensing detections of dwarf galaxies with combined $S/N=38$ and a significant measurement down to \logmass~$=8.0$. We show that the one-halo regime (scales $\lesssim 0.15h^{-1}\rm Mpc$) is insensitive to various systematic and sample selection effects, providing robust halo mass estimates, while the signal in the two-halo regime depends on galaxy color and environment.
These results demonstrate that DESI already enables precise dwarf lensing measurements, and that calibrated photometric samples extend this capability. Together, they pave the way for novel constraints on dwarf galaxy formation and dark matter physics with upcoming surveys like the Vera C. Rubin Observatory's LSST.

\end{abstract}

\section{Introduction}
The Lambda Cold Dark Matter ($\Lambda$CDM) model has had immense success \citep[e.g.,][]{Planck2020} but puzzles remain at small ($<1$ Mpc) scales \citep[e.g.,][]{Bullock2017}. It is unclear whether these challenges reflect a deviation from the cold dark matter model or if they signal inadequacies in our understanding of galaxy evolution.  More broadly, small-scale cosmic structure is poorly constrained and can provide unique information on dark matter physics \citep[e.g.,][]{Boddy2022}.
In this context, 
dwarf galaxies play a key role, especially because they are relatively dark-matter dominated \citep[e.g.,][]{deBlok2010,CollinsRead2022}.

Studies of dwarf galaxies that incorporate baryons into high-resolution simulations
have alleviated some pressures on $\Lambda$CDM, such as the ``missing satellites'' and ``too big to fail'' problems, yet questions remain \citep[e.g.,][]{BrooksZolotov2014, Wetzel2016, Brooks2017, Martin-Alvarez2022}. To distinguish between dark matter scenarios, we must measure the mass profiles for statistical samples of low-mass galaxies \citep{Fry2015,Tollet2016,Kamada2017,Robles2017,Fitts2019,Nadler2021}. At the same time, we need improved constraints on the slope, normalization, and scatter of the low-mass end of the stellar-to-halo mass relation (SHMR). There are also open questions on how properties other than mass correlate to those of the dark matter halo \citep[e.g.,][]{Wechsler2018}.

Many measurements of the SHMR have relied on theoretical assumptions (e.g., halo occupation distribution modeling, \citealt[][]{Yang2012}, or abundance matching, \citealt{ConroyWechsler2009,Behroozi2013}) rather than direct measurements.  Kinematics approaches provide a direct dynamical measurement, but are limited to small samples \citep{Rubin1978, Ferrero2012,Boylan-Kolchin2012,Simon2019, Oh2015,McQuinn2022}. In the era of deep, wide-field photometric surveys, weak galaxy--galaxy lensing (GGL) has become an exciting tool for the detailed study of large samples of galaxies \citep[e.g.,][]{Mandelbaum2006,Leauthaud2012,Mandelbaum2013,Velander2014,vanUitert2016,Zacharegkas2022,Chaurasiya2024}. The halo mass profile measured with gravitational lensing not only provides the average halo mass and satellite fraction of galaxy samples but is also the best way to address the ongoing challenge of understanding the shape of inner halo mass profiles in extragalactic systems \citep{Burkert1995,Bullock2017,Drlica-Wagner2019}.

Detecting the subtle weak lensing signal from dwarf galaxies (defined in this work as galaxies with \logmass~$<9.25$) requires large statistical samples. Recent studies have begun to explore weak lensing in this new mass frontier, demonstrating that such measurements are now within reach \citep{Thornton2024,Chicoine2024}.
In the era of the NSF--DOE Vera C. Rubin Observatory's Legacy Survey of Space and Time \citep[LSST,][]{Ivezic2019}, deeper imaging will enable measurements of lower halo masses ($\rm M_{\rm halo} \lesssim 10^9 \, M_{\odot}$) and smaller scales \citep[$\lesssim 5 h^{-1} \rm kpc$,][]{Drlica-Wagner2019}.
A methodological challenge remains in identifying a sufficiently large sample of dwarf galaxies and characterizing its redshift and stellar mass distribution.
\citet{Tanoglidis2021} have addressed this challenge by focusing on low surface brightness galaxies, while \citet{Luo2024} outline a plan to use medium-band filters for photometric redshifts, and \citet{xSAGA} used a convolutional neural network to identify a large low-mass galaxy sample using only images. While recent studies have taken the first steps in dwarf galaxy lensing, our work delivers the deepest and highest-precision constraints yet, enabled by both spectroscopic and photometric samples.

As we demonstrate, the sample of dwarf galaxies observed by Dark Energy Spectroscopic Instrument \citep[DESI,][]{DESI2016a.Science,DESI2022.KP1.Instr,DESI2023b.KP1.EDR,SurveyOps.Schlafly.2023,Spectro.Pipeline.Guy.2023,Corrector.Miller.2023,FocalPlane.Silber.2023,Snowmass2013.Levi,Poppett_2024} is now large enough to be used as a spectroscopic lens sample. 
However, to increase the signal-to-noise ratio of measurements and take advantage of the continued outpacing of spectroscopic surveys by photometric ones, we also extend the work of \citet[][hereafter T24]{Thornton2024} to select a photometric sample of dwarf galaxies. We use imaging from the Dark Energy Survey \citep[DES,][]{Abbott2018,Abbott2022} and calibrate the stellar masses and redshifts using DESI measurements.
Thanks to the significant increase of the size of the dwarf sample and the precision and accuracy of the calibration, we are able to push to unprecedentedly low halo masses and measure halo mass profiles for subsamples with different galaxy properties.

In Section \ref{sec:data}, we describe the DES, DESI, Kilo-Degree Survey \citep[KiDS,][]{deJong2013}, and Sloan Digital Sky Survey \cite[SDSS,][]{York2000} samples that we use for lenses and sources. 
A subset of the DESI lenses are used as calibration for the DES lenses.
In Section \ref{sec:ggl} we provide an overview of our methodology for galaxy--galaxy lensing. 
We then introduce the lensing measurements for the spectroscopic lenses in Section \ref{sec:desi-lensing}.
Section \ref{sec:sample-selection} presents the self-organizing map \citep[SOM,][]{Kohonen1982,Kohonen1990} technique that we use to characterize and bin the photometric lens sample. We describe our DES lensing profile measurements in Section \ref{sec:des-lensing}, including comparisons to the spectroscopic lenses.
We discuss these results in Section \ref{sec:discussion} and summarize with a focus on future opportunities in Section \ref{sec:conclusion}.

\section{Data}
\label{sec:data}

\begin{figure*}
    \centering
    \includegraphics[width=1\linewidth]{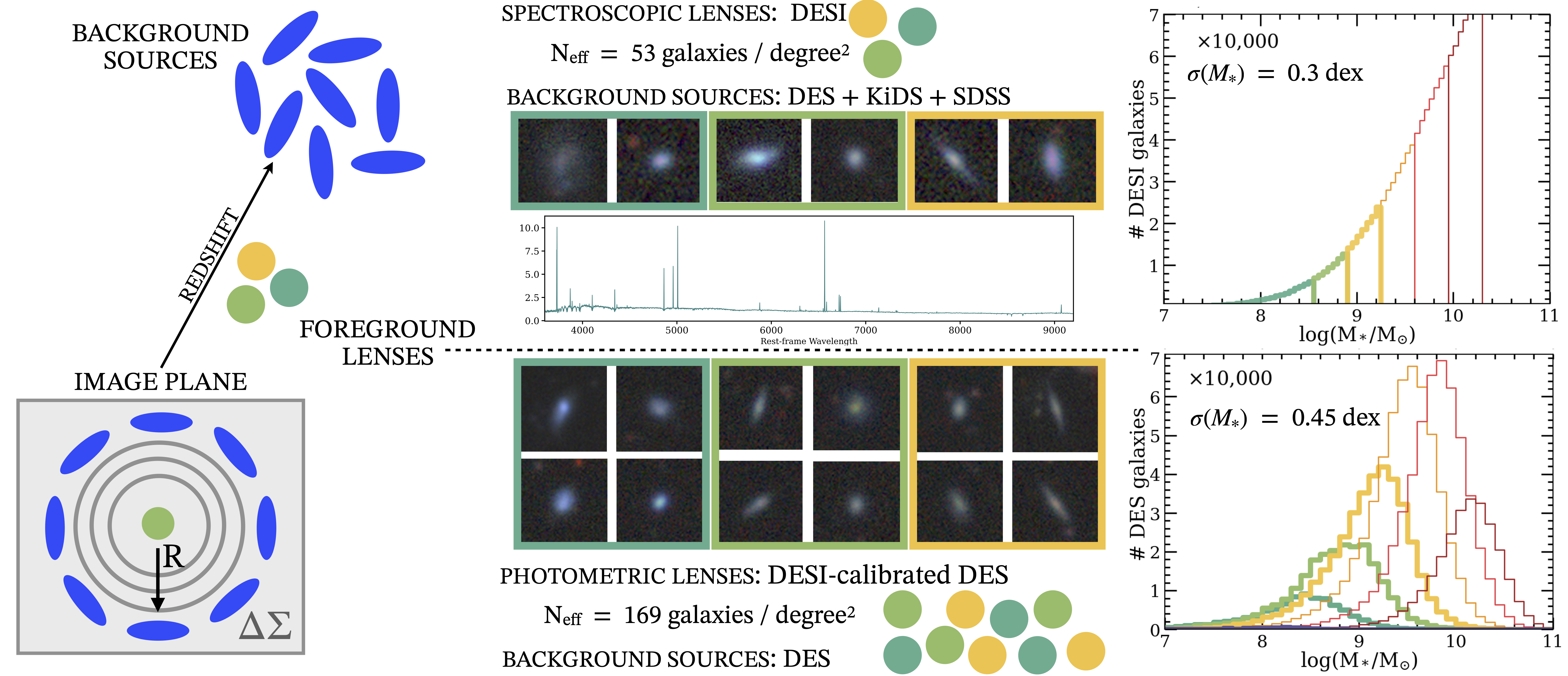}
    \caption{Schematic summarizing the data used for the two analyses in this work. We measure galaxy--galaxy lensing using two samples of foreground lenses, each divided into six stellar mass subsamples, three of which target dwarf galaxies with \logmass~$<$~9.25, indicated as teal, green, and yellow circles in the figure, and with postage stamps of randomly-selected galaxies from each. The two samples are: (1) A spectroscopic sample from DESI (upper panel) with precise stellar masses for each galaxy (with a systematic error of $\sigma(M_*)=0.3$, Section~\ref{sec:desi-data}) allowing for well-defined bin edges (right). The stacked spectrum from the lowest-mass subsample is shown. In the spectroscopic sample, we have
    30,465 dwarf galaxies overlapping the DES footprint with a corresponding density of $\sim$50 galaxies/deg$^2$. This DESI sample is limited to the area that overlaps background lensing data, using sources from DES, KiDS, and SDSS (see Section \ref{sec:sources}). (2) A DES photometric sample, which we select and characterize using DESI (Section~\ref{sec:des-lens-sample}) to achieve a $\sim\times$3 higher number density for dwarf galaxies and a substantially larger dwarf sample of 701,123 galaxies, albeit with wider stellar mass distributions ($\sigma(M_*)=0.45$). This number represents the spread in stellar mass for the SOM cells that contribute to the dwarf galaxy bins (Section~\ref{sec:sample-selection}). For this photometric analysis, we use DES background sources. }
    \label{figsummary}
\end{figure*}
Galaxy--galaxy lensing measures the correlations of the shapes of background ``source'' galaxies with the positions of foreground ``lens'' galaxies. In this section we describe the two lens samples that we study (Sections~\ref{sec:desi-data} and \ref{sec:des-lens-sample}) and the source galaxy samples (Section~\ref{sec:source-data}) that we use for our lensing measurements. Figure~\ref{figsummary} summarizes the use of these samples. The target of this work is the dwarf galaxy regime, but we also analyze samples with \logmass~$>9.25$ as a benchmark for our methodology.

DESI and DES data are central to this analysis.
We use DESI galaxies as lens galaxies and for the spectroscopic calibration sample that characterizes a larger photometric dwarf lens sample from DES. We also use DES data as a background source sample for the lensing measurements.

\subsection{DESI spectroscopic foreground lenses}
\label{sec:desi-data} 
\begin{figure}
    \centering
    \includegraphics[width=1\linewidth]{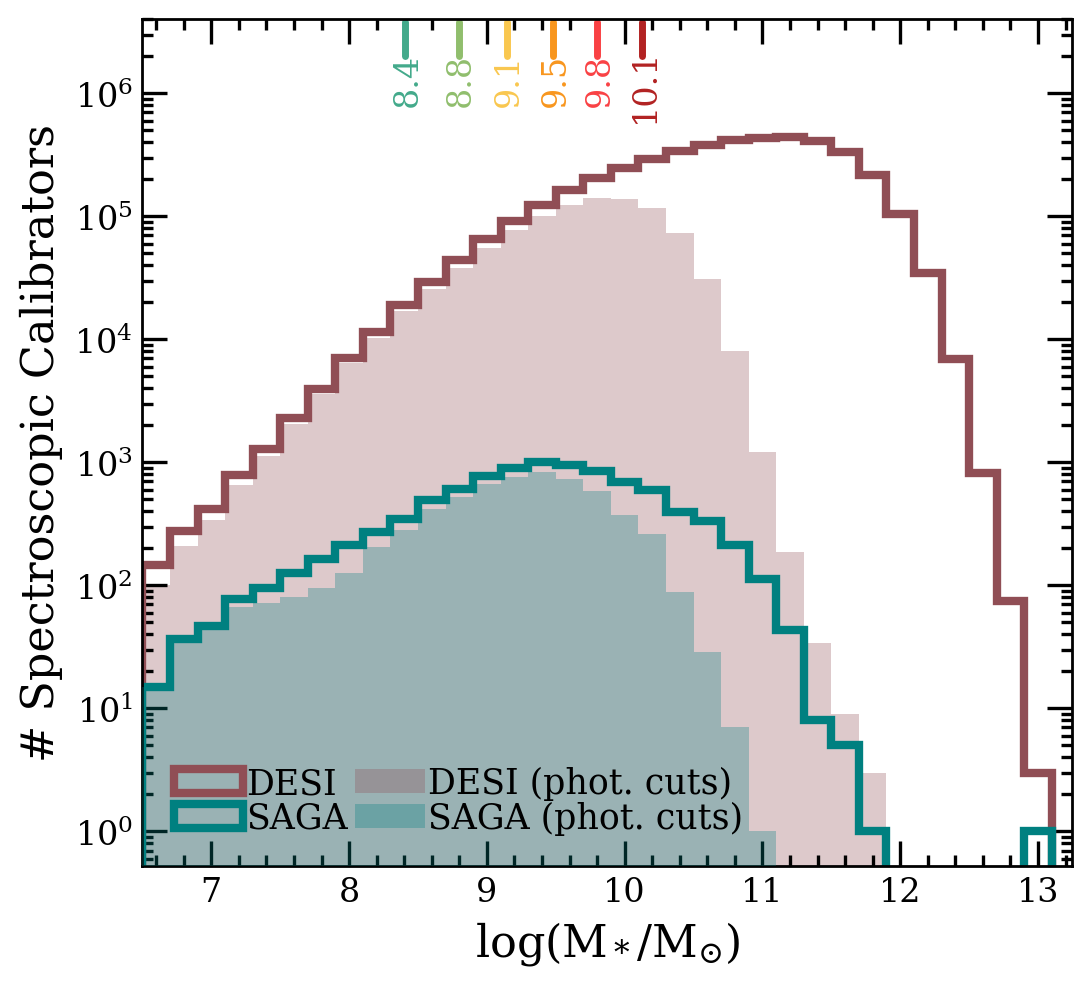}
    \caption{Stellar mass distributions for SAGA's background galaxy sample, as used in \citetalias{Thornton2024} (teal shaded region), and a substantially larger DESI sample used in this work. 
    We use the DESI galaxies both as a spectroscopic lens sample (solid red lines, full sample) and with selection criteria imposed (Equation~\ref{eqn:selection}, shaded red region) to choose a preferentially low-mass sample, which we use to calibrate a larger photometric sample of galaxies from DES. 
    We divide the DESI-calibrated DES galaxies into six stellar mass bins (shown in Figure~\ref{fig:mass_bins}), and we indicate here their median stellar masses at the top.}
    \label{fig:data-summary}
\end{figure}

There are over 250,000 dwarf galaxies and over 1 million galaxies with $9.25<$~\logmass~$<10.3$ in our sample from DESI Data Release (DR) 1.
The average number density of these dwarf galaxies is $\sim$50  
galaxies/$\rm deg^2$. 
In Figure~\ref{fig:data-summary}, we show the mass distribution of the DESI galaxies and highlight the significant sample size improvement. 
We use the DESI Data Release 1 spectra from the BGS \citep{Hahn2023} and LOWZ Dark (\citealt{Darragh-Ford2023}; see also Manwadkar et al. in preparation). 
LOWZ Dark contributes 7\% of our \logmass~$<10.3$ sample but is crucial for filling in the faint/low-redshift parameter space. 
In particular, the LOWZ target selection was designed to be $>95\%$ complete for $z<0.03$ and $r<21$.  

DESI DR1 provides fiber completeness weights for the BGS galaxies included in the large-scale structure (LSS) catalogs \citep{Ross2025}. In galaxy--galaxy lensing estimators, these weights correct for the higher fiber incompleteness in crowded regions \citep[][]{Smith2019,Hahn2023,Lange_2024}.
Fiber completeness weights are not yet available for 24\% of our dwarf (\logmass~$<9.25$) sample and 13\% of our $9.25<$~\logmass~$<10.3$ galaxies.
In Section \ref{sec:comparison} we show that the effect of this missing fraction is unimportant for the current signal-to-noise of our halo mass constraints.

In this work, we use the photometry from the DESI Legacy Imaging Survey \citep{Dey2019} in $g$, $r$, and $z$ bands to determine how lensing signals change with photometric properties.
The imaging is primarily from the Dark Energy Camera Legacy Survey (DECaLS) with supplementation for Dec.~$>32^{\circ}$ from the Beijing-Arizona Sky Survey \citep[BASS,][]{Zou2017} and the Mayall $z$-band Legacy Survey (MzLS). 
To correct for the minor throughput differences between the surveys' bandpasses, we convert to DES magnitudes using the \cite{Dey2019} color transformations, which yield offsets of $<0.05$ mag for $r$ and $z$ and a mean of 0.2 mag for $g$-band. To ensure that we exclude stars and as many false positives as possible, we impose quality cuts\footnote{For consistency, we use the LOWZ requirements \citep{Darragh-Ford2023} for the full sample: $-0.1<g-r<1.5$ and the following in at least two of the $grz$-bands: \texttt{FRACFLUX} $<$ 0.35; \texttt{RCHISQ} $<$ 2; \texttt{SIGMA\_GOOD} $\geq$ 5; \texttt{SIGMA\_GOOD} $\geq$ 30 or \texttt{RCHISQ} $<$ 0.85, where \texttt{SIGMA\_GOOD} is 0 for \texttt{RCHISQ} $>$ 100 and $\texttt{FLUX} \times \sqrt{\texttt{FLUX\_IVAR}}$ otherwise.}. We use the DESI Redrock redshifts \citep[][Bailey et al. in preparation]{Anand2024} and filter out failed redshifts by only including objects with \texttt{ZWARN} $=0$ and \texttt{DELTACHI2} $>40$. 

We estimate the stellar mass for each DESI galaxy following \citetalias{Thornton2024} and the SAGA approach, which is optimized for low-mass galaxies: $\log_{10}(M_*/M_{\odot}) = 1.254 + 1.98(g-r) - 0.4M_r$ with a systematic error of 0.2 dex in the context of SAGA \citep{Mao2021}. 
As we detail in Appendix \ref{sec:stellar-mass-appendix}, we consider three additional sets of DESI stellar mass estimates to check the typical stellar mass uncertainty and to test whether our galaxy--halo connection findings are sensitive to stellar mass method. 
The \cite{delosReyes2024} relation is also tailored to dwarf galaxies and uses redshift and magnitudes. Compared to the estimates using the SAGA approach, these stellar masses are in good agreement, differing by $\sim$0.15 dex. 
The SED$flow$ \citep{Hahn2022} estimates use a neural network trained on PROVABGS \citep{Prova2023} SED models. The average deviation from our default method is close to zero, though there is scatter up to $\sim$0.3 dex. 
Finally, for galaxies with \logmass~$<10.3$, the \texttt{fastspecfit} \citep{Moustakas2023} estimates are offset, on average, by 0.3 dex from the relation used in SAGA.
\cite{delosReyes2024} has compared these methods in the dwarf galaxy regime and shown that these last two approaches can be biased for dwarf galaxies.

For the galaxy--galaxy lensing measurements a random lens catalog is required. We use the BGS DR1 random catalog and sample the redshifts to match our full lens redshift distribution. 

\subsection{DES photometric foreground lenses}
\label{sec:des-lens-sample}
We select a photometric dwarf sample from 5,000~degrees$^2$ of imaging from the Dark Energy Survey DR2 \citep[DES,][]{Abbott2022}. 
First, we select a sample of likely low-mass galaxies by restricting the DES data to $18<r<21$ and imposing the $z<0.03$ completeness cuts defined in \citet{Darragh-Ford2023} (which build upon the selection that SAGA used to prioritize low-redshift galaxies in \citealt{Mao2021}):
\begin{equation}\label{eqn:selection}
\begin{split}
    \mu_{r,\mathrm{eff}} + \sigma_{\mu} - 0.7(r-14) &> 16.8 \\
    g-r - \sigma_{gr} + 0.06(r-14) &< 0.99,
\end{split}
\end{equation}
where all magnitudes are extinction-corrected using \cite{Schlegel1998} and $\sigma_{gr}$ is $\sqrt{\sigma_{g}^2 + \sigma_{r}^2}$. Following \cite{Mao2021}, $\mu_{r,\mathrm{eff}} = r + 2.5\mathrm{log_{10}}(2\pi R_{\mathrm{eff}}^2)$, where $R_{\mathrm{eff}}$ is the effective half-light radius in arcseconds. 
The motivation here is two-fold: the sampling and completeness are then well-defined and the cuts remove galaxies with higher stellar mass. 
These photometric cuts yield a DES \citep{Abbott2022} candidate low-mass sample of 2,104,685 galaxies. 
This sample is 2.8 times larger than the \citetalias{Thornton2024} sample because we use the extended selection criteria for completeness out to $z<0.03$ instead of $z<0.01$ and because we push to $r$-band magnitude of 21 rather than 20.75. 

The second step is to characterize the photometric lens sample. To do so, we improve on the self-organizing map (SOM) methodology presented in \citetalias{Thornton2024}, which we detail in Section~\ref{sec:sample-selection}. This approach requires a spectroscopic calibration sample, which we create by applying the same selection criteria (the $z<0.03$ completeness cuts and $18<r<21$) to the DESI DR1 data. A key advantage of this dataset is the inclusion of $z$-band magnitudes, which are not available for the \citetalias{Thornton2024} SAGA calibration sample. For uniformity, we construct our DES lens catalog using the photometry provided by DR10.1 of the DESI Legacy Imaging Survey, since they apply the same pipeline (\texttt{legacypipe)} to all available DECam data. 
The selection removes nearly all massive galaxies (99.9\% of those with $\rm M_*>10^{11}\,M_{\odot}$) while retaining most low-mass ones (76\% of those with $\rm M_*<10^{10}\,M_{\odot}$).
LOWZ Dark contributes just 9\% of the calibration sample but is responsible for all the galaxies with $r>20.3$. 

The calibration sample is left with 980,269 galaxies, a remarkable increase over the 6,330 used in \citetalias{Thornton2024}. 
We note that this difference is skewed toward higher masses; the multiplicative improvement is 60 for $\rm M_{*}<10^9 \,M_{\odot}$ galaxies. 
The resulting photometric sample achieves a number density of $\sim$170 galaxies/deg$^2$ in the dwarf regime (\logmass$<$9.25), a factor of three improvement over the spectroscopic sample. 

\subsection{Background source galaxies}\label{sec:source-data}
\label{sec:sources}
To compute the lensing signals of the DESI-calibrated DES galaxies, we use DES background source galaxies. For the DESI galaxies, 
we increase the effective overlap area by combining DES sources
 with those from SDSS \citep{York2000} and KiDS \citep[KiDS,][]{deJong2013} using inverse variance weighting \citep[e.g.,][]{Amon2022,Heydenreich2025}. KiDS and DES overlap with DESI in independent patches of the sky, and we find that these two surveys give consistent lensing signals \citep[see also][]{Heydenreich2025}. When including SDSS data, we only use area that is not covered by DES and KiDS.

\textbf{DES source sample:} We leverage the Y3 lensing catalog \citep{Gatti2021}, which uses \texttt{METACALIBRATION} \citep{Huff2017,Sheldon2017} for shear measurements.
The photometric redshift calibration \citep{Myles2021} includes four redshift bins. Following \citetalias{Thornton2024}, we exclude the lowest-redshift source bin when using the DES lens sample, because of the extensive overlap between the lens and source photometric redshift distributions. 
This choice removes 25\% of the 100,204,026 galaxies in the catalog. For the DESI lenses, which have well-defined redshifts, we include DES source galaxies in the lowest-redshift tomographic bin (and similarly, do not impose any redshift selection on the KiDS or SDSS samples). We note that the on-sky overlap with DESI that can be used for lensing is limited to 662~degrees$^2$ \citep{Heydenreich2025} and 129,655 \logmass$<$10.3 lenses.
We obtain the shear bias corrections, $\overline{m}$, and the response factors, $\overline{R}$, from \cite{MacCrann2022} and \cite{Prat2022}, respectively. 

\textbf{KiDS source sample:} We use the gold sample of weak lensing and photometric redshift measurements from the fourth KiDS data release \citep[][]{Kuijken2019,Wright2020,Hildebrandt2021}, hereafter referred to as KiDS-1000. The on-sky overlap area with DESI is 448~degrees$^2$ \citep{Heydenreich2025}, including 208,258 \logmass$<$10.3 lenses. KiDS-1000 includes sources three magnitudes fainter than SDSS. 
We use the \citet[][]{Giblin2021} multiplicative correction.

\textbf{SDSS source sample:} We use the SDSS DR7 weak lensing catalogs, which cover over 11,000 square degrees to a limiting magnitude of 22 \citep{Abazajian2009} The total on-sky overlap area with DESI is 5,459~degrees$^2$ \citep{Heydenreich2025}.
Accounting for the regions that we exclude in favor of DES and KiDS, the area includes 721,311 \logmass$<$10.3 lenses. \cite{Reyes2012} presents the shape catalog, while \cite{Mandelbaum2012} and \cite{Mandelbaum2013} provide the systematic tests and correction factors.  SDSS has $\overline{m}$ and a shear responsivity correction that is a property of the estimator and not of galaxy blending. We use the photometric redshift distributions determined by \cite{Nakajima2012}.

\section{Galaxy--galaxy lensing methodology}
\label{sec:ggl}
We briefly summarize the methodology for the galaxy--galaxy lensing measurements in Section~\ref{sec:lensing-overview}. For more details, we refer the reader to \citet{Amon2022}, \citet{Thornton2024}, and \citet{Heydenreich2025}. The modeling approach to estimate the halo mass from the lensing profiles is discussed in Section~\ref{sec:nfw-modeling}. 

\subsection{Measuring galaxy--galaxy lensing}
\label{sec:lensing-overview}
Foreground galaxies induce percent-level shears on the background source galaxies. Because this shear only occurs tangent to the lens galaxy, the lensing becomes measurable when we stack over a large sample. 
We measure the average two-point position-shear correlation function, $\langle \gamma_{t}(\theta) \rangle$. For a lens galaxy at redshift $z_{\rm l}$, $\langle \gamma_{t}(\theta) \rangle$ can be converted to a physical lensing profile using the comoving distance to the lens, $\chi(z_{\rm l})$, to the source, $\chi(z_{\rm s})$, and between the lens--source pairs, $\chi(z_{\rm l},z_{\rm s})$. 
In particular, the (uncorrected) comoving quantity is the excess surface mass density, $\Delta \Sigma_{\mathrm{uncorr.}}$, defined as a function of the comoving radius, $R=\chi(z_{\rm l})\theta$, from the lens: 
\begin{equation}
\label{eq:uncorrected}
    \Delta \Sigma_{\mathrm{uncorr.}}(R) = \frac{c^2 }{4\pi G (1+z_{\rm l}) }\frac{\chi(z_{\rm s})}{ \chi(z_{\rm l}) \chi(z_{\rm l},z_{\rm s})} \langle \gamma_{t}(\theta) \rangle \, . 
\end{equation}
This quantity represents the difference between the average surface mass density within the radius $R$ and the projected surface density at that $R$.
As detailed in Section 4 of \citetalias{Thornton2024}, this equation can be generalized for calibrated redshift probability distributions. We estimate uncertainties using a Jackknife covariance with 150 regions determined with \textsc{k-means}, although we confirm that it is not sensitive to the number of regions.

Several corrections are necessary. The multiplicative corrections include the response factor ($\overline{R}$) and multiplicative bias correction ($\overline{m}$).
To account for additive systematic uncertainties, we subtract any excess surface density signal found for random positions within the survey footprint with the same redshift distribution as the lenses\footnote{For the spectroscopic lenses, we use the BGS DR1 random catalog. For the photometric lenses, we employ the same random catalogs used in \citetalias{Thornton2024}, which include over 10 million randoms.}, $\Delta\Sigma_{\mathrm{rand.}}$ \citep{Mandelbaum2005,Mandelbaum2013,Singh2017}.
We use the same randoms for the ``boost factor'' correction, $B(R)$, which addresses the overlap between lens and source redshift distributions \citep[e.g.,][]{Sheldon2004,Amon2018}. Putting these corrections together yields
\begin{equation}
    \Delta \Sigma_{\mathrm{corr.}} = \frac{B(R)\Delta \Sigma_{\mathrm{uncorr.}} - \Delta \Sigma_{\mathrm{rand.}}}{\overline{R}(1+\overline{m})} \, .
\end{equation}
As a validation check, we confirm that the ``cross-shear'' $45^{\circ}$ from the tangential shear is consistent with zero. 
Throughout this work, we assume \cite{Planck2020} cosmological parameters and compute $\Delta \Sigma(R)$ in comoving coordinates in units of $h^{-1}$Mpc.

\subsection{Estimating halo mass from a lensing profile}
\label{sec:nfw-modeling}
To estimate halo masses, \citetalias{Thornton2024} fit their lensing signals with both a CDM $N$-body simulation-based model and a Navarro-Frank-White \citep[NFW,][]{NFW1997} profile. 
They found that these two approaches yield consistent halo masses in the inner radial region.
\citetalias{Thornton2024} (see Figure 10 of that work) also demonstrated that the inner region of the lensing profile (i.e., the one-halo regime) provides a good estimate of halo mass and is not significantly impacted by the satellite fraction. 
Motivated by this finding, here we fit an NFW profile to the lensing signals for an initial estimate of the halo masses. We leave a more sophisticated simulation-based approach for future work; employing such methods robustly to our lowest-mass samples requires higher-resolution and larger-volume simulations. Instead, we focus on improving the dwarf selection method, comparing photometric and spectroscopic samples to ensure robustness, and studying the dependence of the signal on galaxy properties beyond stellar mass. 

Informed by the scale at which the NFW and simulation-based models diverge in \citetalias{Thornton2024}, we use a conservative cutoff for the scales included in our NFW model fits ($<0.15 h^{-1}\rm Mpc$), which corresponds to the inclusion of six lensing measurement data points. This choice is further supported by the finding that the sample incompleteness investigated in this work has a statistically significant effect on the two-halo signal but not on smaller scales (Section~\ref{sec:des-lensing}).

Our fitting procedure follows Appendix C2 of \citetalias{Thornton2024}; we outline it briefly here.
Following \cite{Klypin2016}, we define mass as $M_{200} = \frac{4\pi}{3} 200\rho_{\rm cr} R_{200}^3$, where $R_{200}$ is the radius at which the enclosed mean density is 200 times higher than the Universe's critical density.
We construct and fit the NFW profile to the signals using \texttt{pyccl}\footnote{https://github.com/LSSTDESC/CCL} \citep{Chisari2019} and \texttt{emcee} \citep{Foreman-Mackey2013} and use the samples' $\rm 16^{th}$, $\rm 50^{th}$, and $\rm 84^{th}$ percentiles for the plotted halo masses. 
For halo mass and concentration, we use independent priors that are flat in logarithmic space, spanning the ranges $8\times10^{9}$--$2\times10^{12} M_{\odot}$ and $1$--30, respectively.

\begin{figure*}
    \centering
    \includegraphics[width=\textwidth]{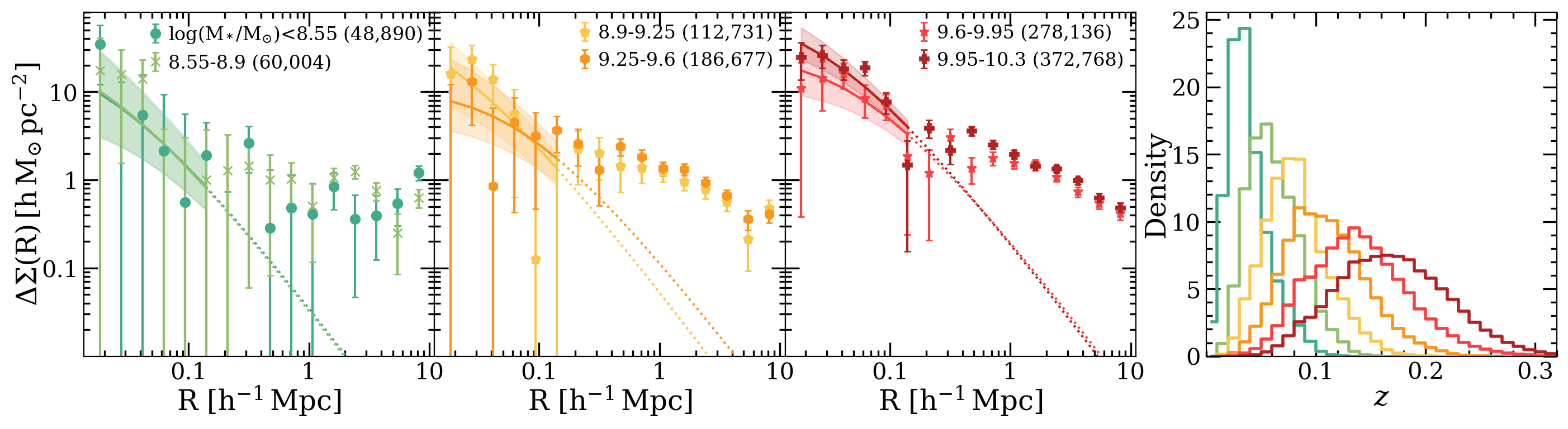}
    \caption{
  Left: Weak lensing halo mass profiles for six stellar mass bins of DESI-fiducial lenses. We detect a signal in all bins: for the dwarf galaxies (\logmass~of $<$8.55, 8.55$-$8.9, 8.9$-$9.25), we find a $S/N$ of 5, 8, and 10, respectively, and a combined $S/N$ of 14.
  The curves represent the best-fit NFW profiles, with the shading showing the one-sigma uncertainty. The solid portion includes the six data points used in the fit. We also note the number of DESI lenses per bin.
  Table~\ref{tab:modeling} includes the resulting halo mass constraints along with the median stellar masses and full profile signal-to-noise ratios.
    Right: Redshift distributions for the six stellar mass bins.
    }
    \label{fig:exact-desi}
\end{figure*}

\begin{deluxetable*}{c cccc cccc}
\tablecaption{Halo mass measurements for the stellar mass selections used throughout this work. For the DESI-fiducial lenses, the stellar mass bin column refers to the range of DESI galaxy stellar masses selected for each lens bin. For the DES case, it refers to the range of mean stellar masses of the SOM cells selected for that bin (see Section \ref{sec:sample-selection}). The reported stellar mass uncertainties include an additional 0.3~dex added in quadrature with the error derived from the $16^{\rm th}$ and $84^{\rm th}$ percentiles. For the halo mass, the median of the posterior from the NFW fit to the inner ($15\,h^{-1}\,\mathrm{kpc}<R<0.15h^{-1}$Mpc) region of the lensing profile is reported with 84th and 16th percentiles providing the statistical uncertainty. We note that the lowest-mass bin reported ({$\log_{10}(M_*/M_{\odot})<$8$^*$}, indicated with $^*$), includes overlapping DES galaxies with the $<$8.55 bin and the halo mass constraint is derived from a fit to the measurement extending to $10\,h^{-1}\,\mathrm{kpc}$, as shown in Figure~\ref{fig:low}. We compute the signal-to-noise ratios for the full measured profiles following \citetalias{Thornton2024}.\label{tab:modeling}}
\tablehead{
\colhead{Stellar mass bin} &
\multicolumn{4}{c}{DESI-fiducial} &
\multicolumn{4}{c}{DES} \\
\colhead{} &
\colhead{N gal} &
\colhead{$\log_{10}(M_*/M_{\odot})$} &
\colhead{$\log_{10}(M_{\rm halo}/M_{\odot})$} &
\colhead{S/N} &
\colhead{N gal} &
\colhead{$\log_{10}(M_*/M_{\odot})$} &
\colhead{$\log_{10}(M_{\rm halo}/M_{\odot})$} &
\colhead{S/N}
}
\startdata
$<8^*$       & $\dots$ & $\dots$ & $\dots$ & $\dots$ & 5,892   & $8.04_{-0.52}^{+0.77}$ & $10.45_{-0.40}^{+0.48}$ & 5  \\
$<$8.55      & 48,890  & $8.28^{+0.36}_{-0.50}$ & $10.68^{+0.40}_{-0.49}$ & 5  & 90,661  & $8.43^{+0.51}_{-0.68}$ & $10.23^{+0.29}_{-0.23}$ & 11 \\
8.55$-$8.9   & 60,004  & $8.75^{+0.32}_{-0.33}$ & $10.68^{+0.36}_{-0.48}$ & 8  & 232,280 & $8.83^{+0.51}_{-0.62}$ & $10.90^{+0.12}_{-0.11}$ & 20 \\
8.9$-$9.25   & 112,731 & $9.10^{+0.32}_{-0.33}$ & $10.98^{+0.21}_{-0.34}$ & 10 & 378,182 & $9.15^{+0.50}_{-0.58}$ & $11.08^{+0.09}_{-0.11}$ & 30 \\
9.25$-$9.6   & 186,677 & $9.45^{+0.32}_{-0.33}$ & $11.02^{+0.31}_{-0.46}$ & 15 & 552,682 & $9.48^{+0.47}_{-0.52}$ & $11.47^{+0.06}_{-0.07}$ & 41 \\
9.6$-$9.95   & 278,136 & $9.79^{+0.32}_{-0.32}$ & $11.43^{+0.16}_{-0.20}$ & 20 & 522,481 & $9.82^{+0.44}_{-0.49}$ & $11.52^{+0.06}_{-0.07}$ & 43 \\
9.95$-$10.3  & 372,768 & $10.14^{+0.32}_{-0.32}$ & $11.55^{+0.09}_{-0.11}$ & 26 & 260,738 & $10.14^{+0.42}_{-0.48}$ & $11.72^{+0.06}_{-0.07}$ & 40 \\
\enddata
\end{deluxetable*}

\section{DESI dwarf lensing measurements}
\label{sec:desi-lensing}
We now present our DESI lensing measurements, referred to as DESI-fiducial. Although the central goal of this work is to measure the lensing profile of dwarf galaxies, we also study the higher-mass regime as a benchmark and to constrain the SHMR over a wider mass range. 
Throughout this work, we use six bins spaced by 0.35 dex in \logmass, as given in Table~\ref{tab:modeling}. Three of these bins include dwarf galaxies (here, \logmass~$<9.25$) and the higher-mass bins range from  \logmass~$=9.25$--$10.3$.
We choose these bin definitions to enable broad SHMR coverage and high signal-to-noise lensing measurements.

For each mass bin, we use 16 logarithmically spaced radial bins spanning from $R=15$ $h^{-1}$kpc to 10 $h^{-1}$Mpc around the lens galaxies. The lower radial bin is defined such that the signal is non-zero for most bins: lower $R$-values yield signals consistent with zero for the higher-mass bins. 
In Section~\ref{sec:how-low}, we extend to 10 $h^{-1}$kpc for the lowest-mass photometric lenses. 
We find that the halo mass constraints are consistent regardless of the binning choices.

For each DESI-fiducial stellar mass bin, we compute the galaxy--galaxy lensing measurements using the \textsc{dsigma} pipeline \citep{Lange2022}. We measure the lensing signal using background galaxies from DES, KiDS, and SDSS (limiting SDSS to the non-overlapping survey area). As the signals from the three source surveys are consistent, we combine them using inverse-variance stacking \citep[e.g.,][]{Amon2023, Heydenreich2025}. 
We note that random subtraction has a minimal effect for all the lensing signals presented in this work. 

The stacked DESI-fiducial $\Delta\Sigma(R)$ measurements and redshift distributions are shown in Figure~\ref{fig:exact-desi}, along with the best-fitting NFW model to the inner regions of the measurements and the uncertainty in the fit. For all mass bins, there is a significant signal detected. Following \citetalias{Thornton2024}, we compute the signal-to-noise ($S/N$) ratios of the lensing measurements with a $\chi^2$ comparison to a null signal.
We include these values in Table~\ref{tab:modeling}.
For the bins in the dwarf regime (\logmass~of $<$8.55, 8.55$-$8.9, 8.9$-$9.25), we find a $S/N$ of 5, 8, and 10, respectively. 
Combining these dwarf galaxy bins, we find a $S/N$ of 14.
As expected, we find that higher-mass stellar mass bins have higher fitted halo masses (Table~\ref{tab:modeling}).

\section{Building a photometric dwarf galaxy sample}
\label{sec:sample-selection}
The DESI-fiducial lensing measurements present a promising future for studying dwarf halo mass profiles with a spectroscopic sample, especially as overlapping lensing surveys improve. However, the substantially larger imaging volume that will soon be available from LSST \citep[][]{Drlica-Wagner2019}, Euclid \citep{Laureijs2011}, and the Nancy Grace Roman Telescope \citep{Spergel2015} motivates the need for methods to overcome the challenges to selecting and characterizing a larger photometric dwarf sample. These imaging surveys have the potential for higher signal-to-noise and measurements at smaller scales and lower masses, thereby yielding novel constraints on both cosmological and galaxy evolution questions.

Section~\ref{sec:des-lens-sample} describes the photometric selection criteria for the sample of likely low-mass galaxies. To characterize this photometric lens sample, we follow the SOM framework demonstrated by \citetalias{Thornton2024}. The SOM is an unsupervised machine learning technique that leverages competitive learning for dimension reduction \citep{Kohonen1982,Kohonen1990}. The SOM groups galaxies into ``cells'' with similar photometric properties. 
By adding the spectroscopic calibration sample to the trained SOM, we determine the empirical stellar mass and redshift distribution for each cell, which we adopt as the distribution for the photometric sample in that same cell.
We then use this information to group cells into stellar mass bins. SOMs are well-established as useful for the photometric redshifts employed by precise cosmological analyses \citep{Masters2015,Buchs2019,Alarcon2020,Wright2020,Hildebrandt2021,Myles2021,Giannini2022,Sanchez2023}. 
By also using them to calibrate stellar masses, we are taking an analogous approach. 

Here, we outline our use of the SOM to select dwarf galaxies from DES imaging. We refer the reader to \citetalias{Thornton2024} and focus here on this work's advancements, many of which are enabled by the significantly larger calibration sample. In Table~\ref{tab:som-summary}, we outline how each change increases the number of \logmass~$<9$ galaxies and changes the mean $\sigma_{\log_{10}(M_*/M_{\odot})}$ across the SOM cells.
\begin{table}
\centering
\setlength{\tabcolsep}{2pt}
\begin{tabular}{cccccc}
\hline
\hline
$z$-cut & $r_{\rm max}$ & \# Cells & Properties & \# & $\rm \sigma_{log_{10}(M*/M_{\odot})}$ \\
\hline
$0.01$&20.75&256&$r,g-r,\mu_{r}$&172,968&0.49\\
$\boldsymbol{0.03}$&20.75&256&$r,g-r,\mu_{r}$&264,596&0.48\\
$0.03$&\textbf{21}&256&$r,g-r,\mu_{r}$&354,024&0.49\\
$0.03$&21&\textbf{1,024}&$r,g-r,\mu_{r}$&387,728&0.45\\
$0.03$&21&1,024&\textbf{+scaling}&417,729&0.41\\
$0.03$&21&1,024& +scaling+$\boldsymbol{r-z}$&422,360&0.39\\
\hline
\end{tabular}
\caption{
Summary of the improvement in the selection of dwarf galaxies from each change to the SOM, quantified by number of galaxies in cells with $\langle$\logmass$\rangle<9$, \#, and the mean stellar mass dispersion across all cells,  $\rm \sigma_{log_{10}(M*/M_{\odot})}$.
In each row, we bold the one change made relative to the row above.
The top row uses the same photometric completeness cuts ($z$-cut), upper magnitude limit ($r_{\rm max}$), number of SOM cells (\# Cells) and properties used to train the SOM (Properties) as \citet{Thornton2024} but with updated DES photometry and DESI calibration data. The bottom row describes the selection methodology used in this work, which includes loosened photometric cuts, more SOM cells, an additional color ($r$--$z$) in SOM training, and property scaling.}
\label{tab:som-summary}
\end{table}

\begin{figure*}
    \centering
    \includegraphics[width=\textwidth]{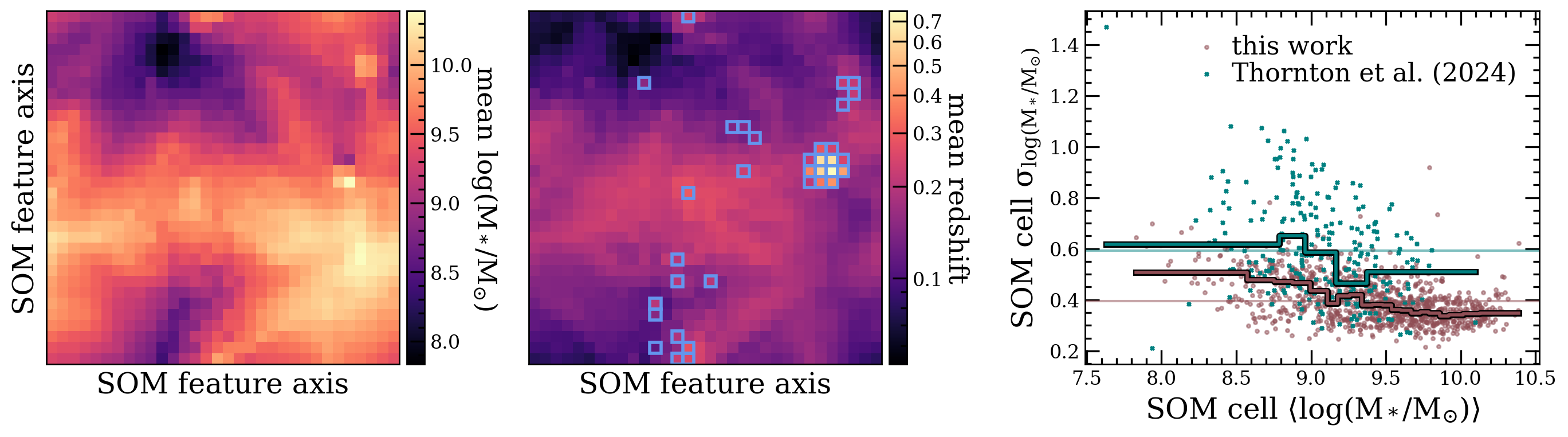}
    \caption{Left: Distributions of mean stellar mass and redshift for DESI calibration sample galaxies within the SOM that is used for calibrating DES lenses. 
    The blue squares in the redshift panel highlight SOM cells dropped because of $\gtrsim0.05$ dex impacts on the mean stellar masses from the $z>0.5$ galaxies (see Section \ref{sec:stellar-mass-appendix}). This cut removes $<3$\% of the DES galaxies.
    Although the SOM is not directly trained to separate galaxies of different stellar masses, there is significant ($>2$ dex) variation of mean stellar mass across the cells.
    Right: $\rm \sigma_{log_{10}(M*/M_{\odot})}$ vs. $\langle$\logmass$\rangle$ for SOM cells in this work (pink) and in \citetalias{Thornton2024} (teal). The mean standard deviation has dropped from 0.58 to 0.39, as indicated by the horizontal lines.
    The thick lines mark the median standard deviation in bins that include 50 SOM cells.
    Compared to \citetalias{Thornton2024}, we can extract more stellar mass bins from the SOM because of the increased number of cells and the decreased $\rm \sigma_{log_{10}(M*/M_{\odot})}$. 
    }
    \label{fig:som-summary}
\end{figure*}

\subsection{Characterizing the photometric dwarf galaxies}
We train the SOM using the DES sample (following $z<0.03$ completeness and $18<r<21$ cuts) and then assign the corresponding DESI calibration sample. Because of the increased size of both the DES and DESI sample relative to those used by \citetalias{Thornton2024}, we use 1,024 rather than 256 cells and four ($r$, $g$--$r$, $r$--$z$, and $\mu_r$) rather than three properties in training.
Our cell number increase is conservative, improving the stellar mass binning while also increasing the number of calibrators per cell: the median number of calibrator galaxies per cell is 425 and each cell has at least 20 calibrators.

We also now scale each of the DES and DESI properties ($P$) using the DES mean ($\mu$) and standard deviation ($\sigma$) according to $(P - \mu)/\sigma$.
This approach prevents arbitrary weighting arising from differences in the photometric properties’ ranges.
For instance, because unscaled surface brightness values range by a factor of $\sim$10, overall vector distances in SOM training are more affected by surface brightness than by the colors. Once we have scaled the properties, the distribution of SOM cell property standard deviations are equivalent---each property is equally weighted in the training procedure. 

It may be beneficial to incorporate further weighting to account for the different stellar mass (or redshift) information content among the properties. In particular, we find that down-weighting surface brightness drives down the typical SOM cell mass dispersion. However, by testing the SOM procedure with only DESI galaxies, we see that additional weighting can introduce a systematic shift on the calibrated stellar masses. If the training and calibration samples have mismatched surface brightness distributions---as they do with the true DES and DESI samples---dropping (or drastically down-weighting) surface brightness leads to cell mean masses that are systematically offset by $\sim$0.05 dex. Because of the difficulty in making this test truly analogous and the minimal potential benefit of weighting, we simply use scaled, equally weighted properties. 

We show the SOM in Figure~\ref{fig:som-summary} with the cells colored by mean stellar mass (left) and redshift (middle).  We see that the SOM generally clusters cells of similar stellar masses, despite being trained only on photometric properties, and can identify low-mass (and low-redshift) galaxies. This behavior provides further support for this approach. 
We exclude cells with uncertainties of $\gtrsim\,$0.05 dex on the mean stellar mass because of the absence of a stellar mass estimate for the $z>0.5$ galaxies (see Section \ref{sec:stellar-mass-appendix}). These dropped cells (blue outline in Figure~\ref{fig:som-summary}) are relatively high mass and include less than $3\%$ of the DES galaxies. 

Combining the effects of our substantially larger samples and our SOM improvements, the mean stellar mass dispersion
per cell decreases from 0.58 in \citetalias{Thornton2024} to 0.39 in this work. In the right panel of Figure~\ref{fig:som-summary}, we show this dispersion as a function of the mean stellar mass of the SOM cell.
These improvements allow us to increase the number of stellar mass bins studied and further divide these bins based on other galaxy properties. 

While more investigation is necessary to confirm whether the SOM is the optimal method, it is likely that the limiting factor is not the methodology but the fact that the photometry provides limited information with which to calibrate stellar mass and redshift. Namely, for much of this photometric parameter space, there is insufficient information to cleanly distinguish between close, low-mass galaxies and more distant, higher-mass galaxies. It is possible that techniques based directly on images may provide more information to split photometric samples, and the photometric information per galaxy may also increase with future surveys like LSST, which will be deeper and include more photometric bands.

\subsection{Calibrated stellar mass distributions}
\label{sec:mass_bins}
\begin{figure*}
    \centering
    \includegraphics[width=\textwidth]{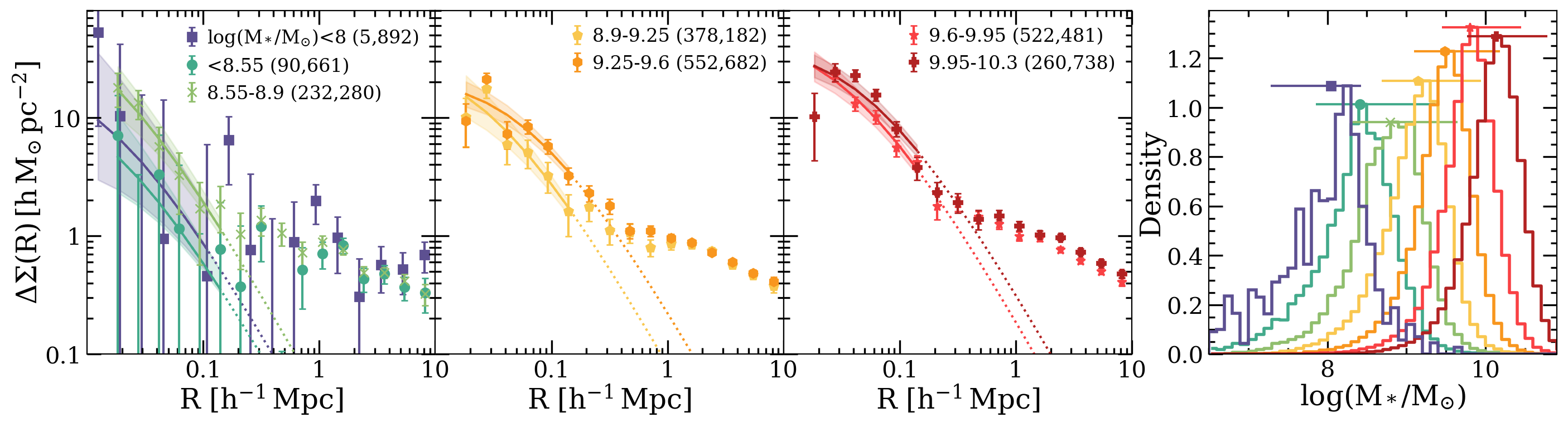}
    \caption{
    Halo mass profiles for six stellar mass bins of DES photometric lens samples and corresponding NFW fits. We include the signal for an additional mass bin, pushing to even lower masses at \logmass~$<8$.
    The stellar mass cutoffs in the legend refer to the mean stellar mass of the SOM cells that enter the bin, defined using the DESI calibrators within that cell. We also note the number of DES lenses per bin. 
    The corresponding calibrated stellar mass distributions are shown in the right panel.
    The points and error bars in the stellar mass distribution panel represent the $16^{\rm th}$, $50^{\rm th}$, and $84^{\rm th}$ percentiles.
    Despite the overlap between the stellar mass distributions, the amplitude of the excess surface density measurements is generally increasing with stellar mass.
    }
    \label{fig:mass_bins}
\end{figure*}
We characterize the DES galaxies in each of the 1,024 SOM cells by adopting the stellar mass and redshift distributions from the DESI galaxies within that cell.
We put the galaxies from each cell into the stellar mass bin (Table \ref{tab:modeling}) whose \logmass~range includes the cell's mean stellar mass. We include in Table \ref{tab:modeling} the resulting number of DES galaxies in each bin.
In the right panel of Figure~\ref{fig:mass_bins}, we show the calibrated stellar mass distributions of the DES galaxies. 
To obtain the distribution for a stellar mass bin, we perform a weighted sum of the distributions from all constituent SOM cells, where the weight for each cell is the ratio of the number of DES galaxies to the number of DESI galaxies in that cell.
We list in Table~\ref{tab:modeling} the corresponding median stellar masses and one-sigma bounds. 

Although the stellar mass distributions are narrower than those presented in \citetalias{Thornton2024} (Appendix \ref{sec:pilot-comparison-appendix}), there is still considerable overlap between the stellar mass bins because of degeneracies between stellar mass and the photometric properties used in the SOM. 
The distributions indicate that 98\%, 91\%, and 67\% of the lenses in the $<8.55$, 8.55$-$8.9, and 8.9$-$9.25 bins have \logmass~$<9.25$.
However, the galaxies within the overlapping region of these stellar mass distributions are systematically different.
For instance, of two galaxies with the exact same stellar mass, the bluer one is more likely to be placed in a lower-mass bin because its color is more typical of a lower-mass population.
In other words, the SOM introduces significant bias (more than straightforward photometric cuts do) because it identifies the low-hanging fruit---parts of photometric property parameter space that tend to be low mass or tend to be high mass. Although we are able to isolate many SOM cells with mean stellar masses in the dwarf regime, a significant fraction of dwarf galaxies are in SOM cells with higher mean stellar masses. We revisit this bias in Section \ref{sec:comparison}.

\section{Photometric dwarf galaxy lensing} 
\label{sec:des-lensing}
We now present the lensing profiles measured for the six DES photometric lens samples.
The primary measurements are shown in Section~\ref{sec:photomeas}. In Section~\ref{sec:comparison}, as a crucial test of this approach, we compare them to the corresponding spectroscopic lens profiles presented in Section~\ref{sec:desi-lensing} (DESI-fiducial). We find consistent halo masses but some inconsistent two-halo signals. This is likely due to the differing photometric distributions and the DESI targeting approach, which we cannot fully correct for without
fiber completeness weights.
We present the resulting SHMR in Section~\ref{sec:shmr} and analyze environmental effects in Section~\ref{sec:environment}.
By investigating the lensing profiles' dependency on galaxy properties beyond stellar mass (Section~\ref{sec:DES_prop_splits}), we find further evidence for both the robustness of the halo mass constraints and the sensitivity of the larger scales to these sample selection effects.

\subsection{Photometric dwarf galaxy lensing measurements} 
\label{sec:photomeas}
To calculate the galaxy--galaxy lensing for DES lenses and background sources, we use a pipeline built around TreeCorr\footnote{https://rmjarvis.github.io/TreeCorr \\ We do not use the \textsc{dsigma} pipeline here as the photometric lens sample has a redshift distribution for each stellar mass bin, rather than per-galaxy redshifts.} \citep{Jarvis2004}. 
We separately measure the signal for each of the three DES source tomographic bins and stack the signals using an inverse-variance average \citep[e.g.,][]{Amon2023}. 

In the left panels of Figure~\ref{fig:mass_bins}, we show the lensing profiles for the six stellar mass bins defined in Section \ref{sec:mass_bins}. The stellar mass distributions are shown in the right panel.
We measure signal-to-noise ratios of 11, 20, and 30 in the \logmass~$<8.55$, 8.55$–$8.9 and  8.9$–$9.25 bins, respectively, and a combined $S/N$ of 38 for these dwarf mass bins. These values represent a substantial ($\sim$$\times2-3$) improvement over the corresponding DESI $S/N$ of 5, 8, and 10 (Table~\ref{tab:modeling}). This statistical gain 
both validates our approach and enables measurement of an even lower-mass bin (\logmass~$<8$). 

Figure~\ref{fig:mass_bins} also shows the best-fit NFW profile for each measurement and its one-sigma uncertainty. As expected, we find an increasing amplitude in the profile (and increasing fitted halo mass) with increasing stellar mass selection. The stellar mass--halo mass relation is discussed further in Section \ref{sec:shmr}.

\subsection{Spectroscopic vs. photometric lensing profiles}
\label{sec:comparison}
\begin{figure*}
    \centering
    \includegraphics[width=1\linewidth]{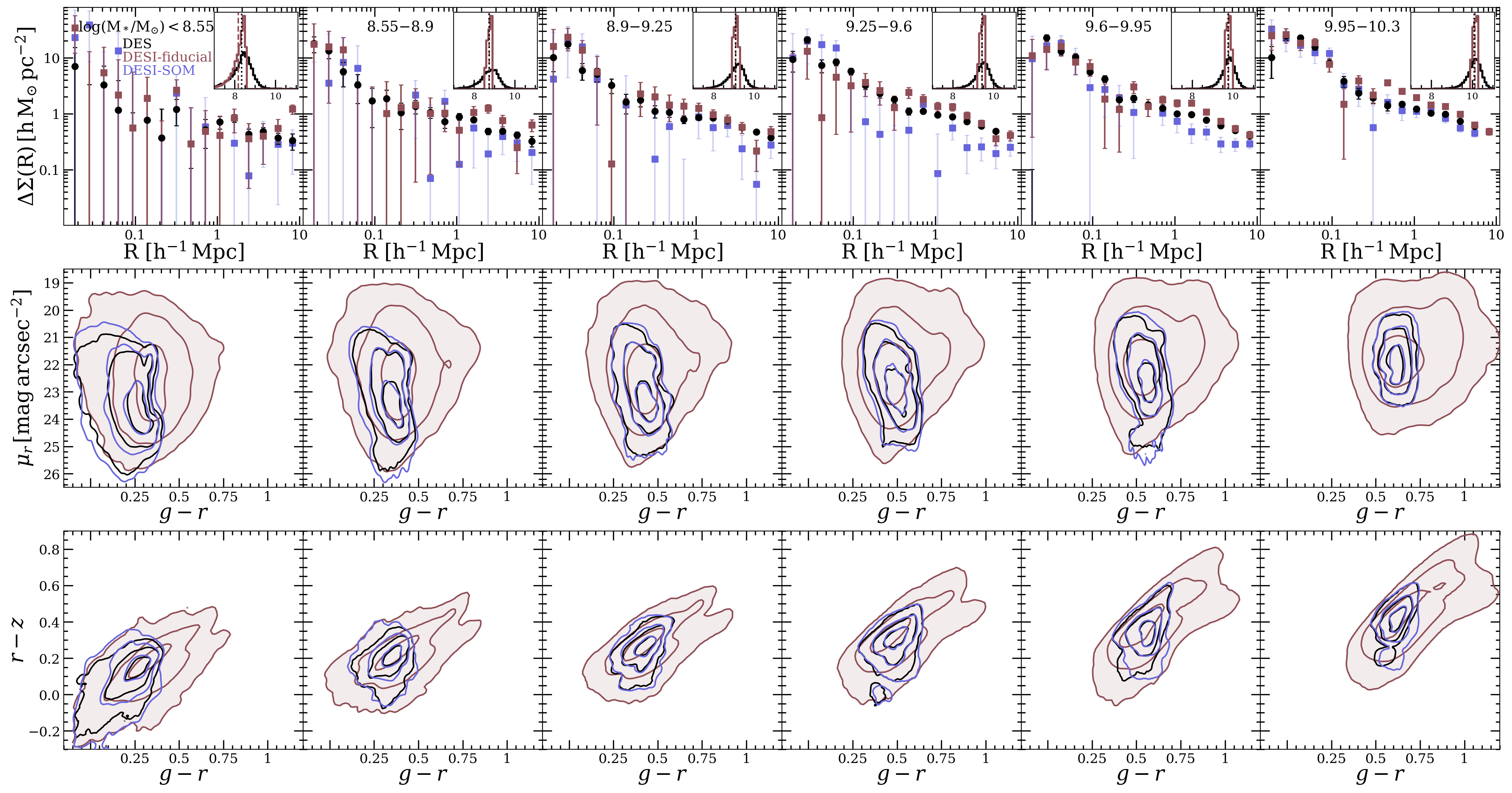}
    \caption{Comparison of photometric and spectroscopic lensing and galaxy properties. Upper row: Comparison of DES galaxy--galaxy lensing mass profiles (black) in the six stellar mass bins to the corresponding DESI-fiducial (red) and DESI-SOM (blue) lensing profiles. 
    The stellar mass distributions are inset, with vertical lines marking the means. The DES distributions are wider but the DES and DESI-fiducial mean stellar masses differ by $\sim0.15$ dex in the lowest-mass bin and $<0.05$ dex in the other five stellar mass bins.
    Middle row: Surface brightness--color ($g$--$r$) distributions for the corresponding selections (six stellar mass bins from lowest to highest), highlighting the effects of the photometric cuts and SOM calibration in the incompleteness of the DES lens sample.
    Lower rows: Corresponding color--color distributions. For most bins, the DES and DESI-SOM distributions are narrower but they share roughly the same central values as the DESI-fiducial ones. However, in the lowest-mass bin, the photometric lenses show a clear bias towards blue values.
    }
    \label{fig:exact_comparison}
\end{figure*}

As a check of the robustness of our methodology to select and characterize a photometric dwarf galaxy sample, we compare the galaxy--galaxy lensing signals computed with the photometric DESI-calibrated DES lenses
to those measured with the DESI lenses (``DESI-fiducial,'' Section \ref{sec:desi-lensing}). 
We plot this comparison for the six stellar mass bins in Figure~\ref{fig:exact_comparison}, along with the stellar mass (inset), surface brightness, and color distributions. 
The DES distribution is biased, especially in color and surface brightness: the SOM excludes redder and higher-surface-brightness galaxies, which tend to be at higher redshift.
These property distributions highlight the incompleteness introduced by the initial photometric cuts (in color, magnitude, and surface brightness) and by the SOM selection.
Despite the substantial differences in the photometric properties and the sample completeness, we see consistency in the inner regions. 

This consistency is clearly quantified by the halo mass constraints from the NFW fits (Table~\ref{tab:modeling}). For the stellar mass uncertainties, we use the $\rm 16^{th}$, $\rm 50^{th}$, and $\rm 84^{th}$ percentiles from the default approach (Section \ref{sec:stellar-mass-appendix}) and add 0.3 dex of systematic error based on the disagreement between the four stellar mass methods investigated (Appendix \ref{sec:stellar-mass-appendix}).

The halo mass consistency is promising, but this comparison makes it difficult to understand what drives the aspects of the profiles that are inconsistent.
For a more direct comparison, we use the precise set of DESI galaxies that fall within the same SOM cells as the lens sample for a given stellar mass bin. That is, we consider only the DESI spectra that match the DES photometric lenses in terms of their surface brightness, magnitude, and color. We refer to this sample as ``DESI-SOM.'' 
Nevertheless, we do not expect completely consistent signals between DESI-SOM and DES, since the ratio of the number of DESI and DES galaxies varies across SOM cells. In an attempt to account for this variation, we weight DESI galaxies in each cell by the product of the fiber completeness weight (where available) and the ratio of the number of DES and DESI galaxies in that cell. As a result, as shown by the contours in Figure \ref{fig:exact_comparison}, the photometric distributions for a given stellar mass bin in these samples are nearly identical, and, unlike in the ``DESI-fiducial'' case, the stellar mass distributions match.

In Figure~\ref{fig:exact_comparison}, we also show these DESI-SOM profiles, which have significantly lower $S/N$. 
We find that the DESI-SOM signals are consistent with the DESI-fiducial and DES signals in the one-halo regime and have consistent corresponding halo mass constraints. However, they tend to have a lower amplitude in the two-halo regime. The inclusion of the DESI fiber completeness weights drives up the two-halo term, since they effectively up-weight crowded regions, which have higher satellite fractions. Thus, this discrepancy between the DES and DESI-SOM signals may simply reflect the current lack of weights for the LOWZ galaxies and for BGS galaxies excluded from the LSS catalogs, which together account for 15\% of the lenses.

\subsection{Stellar mass--halo mass relation}\label{sec:shmr}
\begin{figure*}
    \centering
    \includegraphics[width=2\columnwidth]{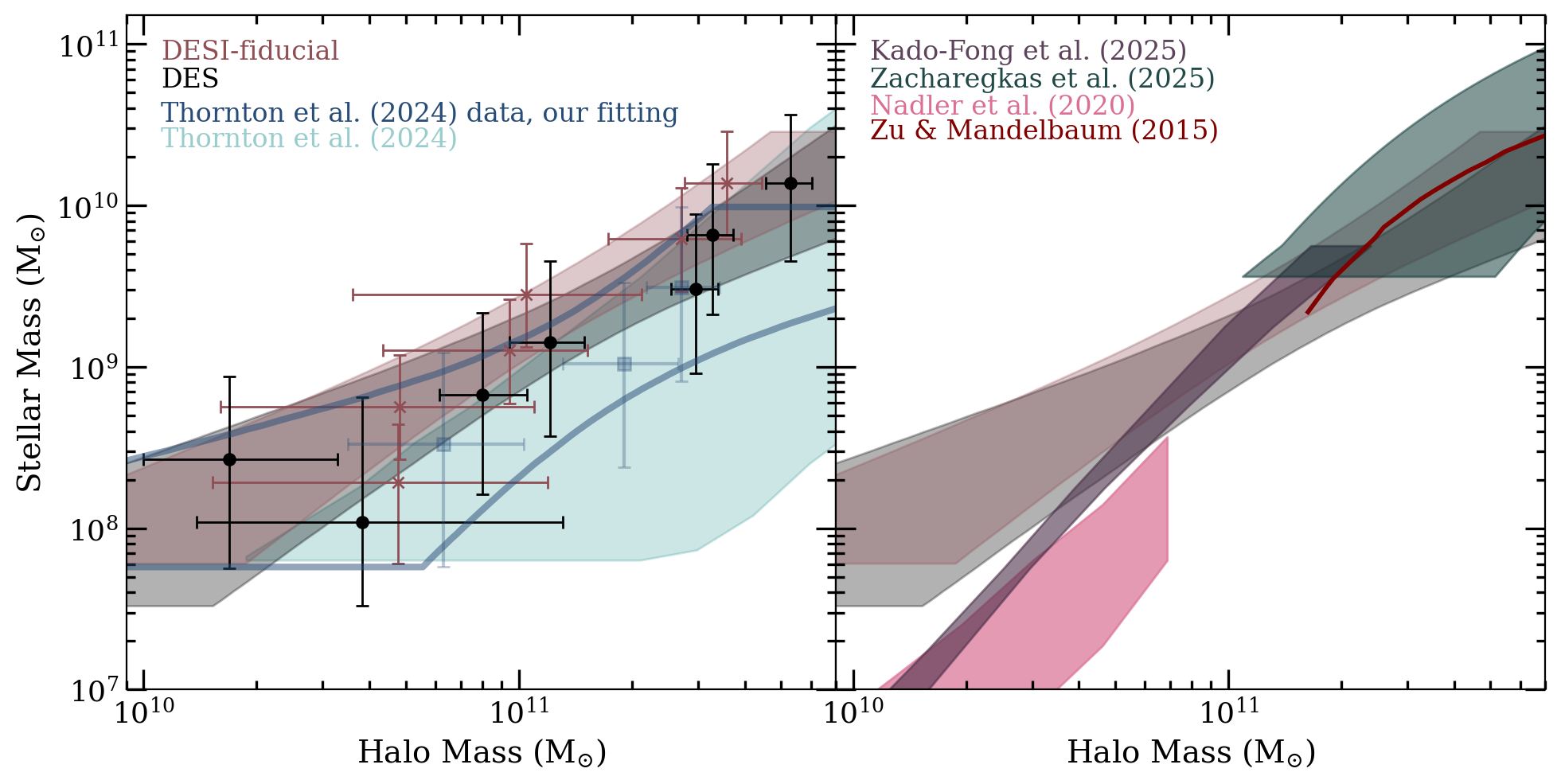}
    \caption{
    Left:
    Stellar vs. halo mass for the six stellar mass bins using the DES lenses (black) and the DESI (pink) spectroscopic lenses from the DESI-fiducial method (Section \ref{sec:desi-lensing}). 
    The black point with the lowest median stellar mass represents the additional \logmass~$<8$ bin.
    For these preliminary halo masses, we fit an NFW to the small scales of the lensing signals ($<0.15 h^{-1} \rm Mpc$; Section \ref{sec:nfw-modeling}).
    The corresponding shaded regions represent the one-sigma uncertainty on the best-fit line in \logmass$-\mathrm{log_{10}}(M_{\rm halo}/M_{\odot})$ space.
    As we discuss in Appendix \ref{sec:pilot-comparison-appendix}, the \citetalias{Thornton2024} constraints (teal) are consistent but systematically offset to higher halo mass for a given stellar mass. However, the magnitude of this effect decreases
    when we apply our NFW and line-fitting approaches to the three plotted \citetalias{Thornton2024} bins (blue).
    Right: 
    Comparison between our DES and DESI-fiducial constraints and those presented in \cite{Zu2015}, \cite{Nadler2020}, \cite{Zacharegkas2025}, and Kado-Fong et al. (2025). We note that the Kado-Fong et al. (2025) results are based on abundance matching assumptions rather than direct mass measurements, and the \cite{Nadler2020} results are based only on satellites of the Milky Way.
    } 
    \label{fig:SHMR}
\end{figure*}

The consistency between the DES and DESI-fiducial mass profiles in the one-halo regime demonstrates the robustness of our inferred halo mass measurements. We now consider the physical implications of the profile strengths as a function of stellar mass.
Using our first constraints in this mass regime (Section \ref{sec:nfw-modeling}), we obtain novel constraints on stellar vs. halo mass, presented in Figure~\ref{fig:SHMR}.

The DESI (pink) and DES (black) data points are shown on the left panel, spanning two orders of magnitude in halo mass, 10~$<\mathrm{log_{10}}(M_{\rm halo}/M_{\odot})<$~12.
Using \texttt{emcee}, we fit lines in \logmass$-\mathrm{log_{10}}(M_{\rm halo}/M_{\odot})$ space and include shading in Figure~\ref{fig:SHMR} for the one-sigma uncertainty on these fits. The corresponding median power-laws for DESI-fiducial and DES lenses are:
\begin{align}
\textrm{DESI}: M_* &=1.756\times10^9(M_{\rm halo}/10^{11})^{1.329}    \\
\textrm{DES}: M_* &=1.277\times10^9(M_{\rm halo}/10^{11})^{1.233}\, .
\end{align} These two relations are in excellent agreement with each other. 

The left panel also includes the three points from \citetalias{Thornton2024} (blue squares), the resulting reported constraints (teal), and the one-sigma uncertainty on the power-law using the same fitting approach described above (blue). We see consistency, although the halo masses are systematically higher for fixed stellar mass. We discuss this discrepancy further in Appendix \ref{sec:pilot-comparison-appendix} and find that the DESI calibration systematically assigns higher mean redshifts and stellar masses to SOM cells than the SAGA calibration from \citetalias{Thornton2024}.

We broadly see agreement with other galaxy--galaxy lensing studies \citep{Zu2015,Zacharegkas2025} as well as with independent constraints that use Milky Way Satellites \citep{Nadler2020} and the abundance matching based on the stellar mass function derived by the SAGAbg sample (Kado-Fong et al. 2025). We note that the  \citet{Nadler2020} and Kado-Fong et al. (2025) measurements both imply somewhat steeper dependence of stellar mass on halo mass. This is an intriguing discrepancy that we will investigate further in future work.
These results should already inform galaxy evolution models, which can predict a wide range of results in this regime. They are likely not yet informative enough to constrain viable dark matter models; this will require pushing the measurements to smaller scales and/or lower masses.

\begin{figure*}
    \centering
    \includegraphics[width=2.1\columnwidth]{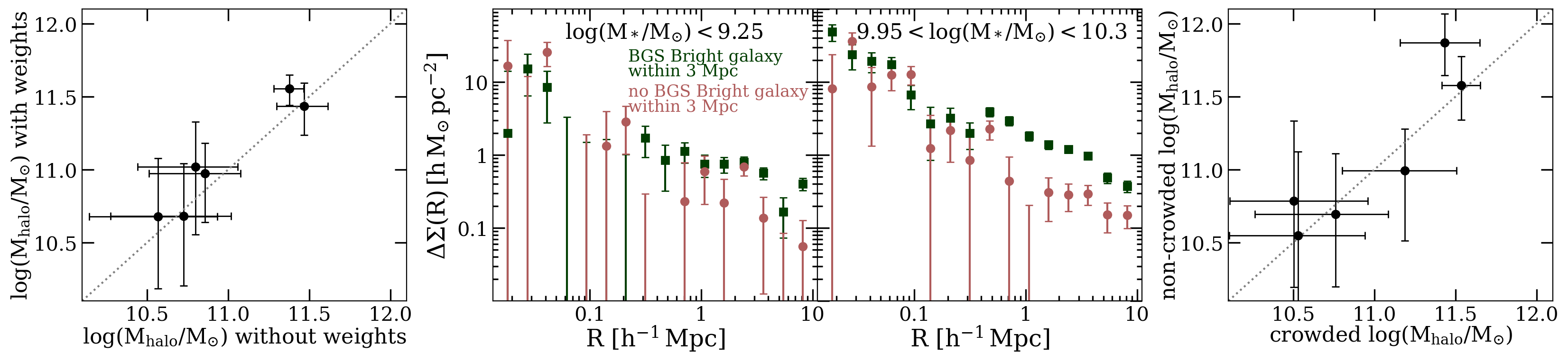}
    \caption{Environmental effects for DESI-fiducial lenses.
    Left: Comparison of halo masses fit to DESI-fiducial signals computed with and without the fiber completeness weights. The highest-mass bin has a disagreement of $>1\sigma$.
    Middle panels:
    DESI lens signals in the combined dwarf ($<$9.25) and $9.95$--$10.3$ ``DESI-fiducial'' stellar mass bins for galaxies with and without a BGS Bright galaxy within 3 Mpc \citep[][]{Hahn2023}. 
    The signals are consistent in the one-halo regime but diverge at larger scales, showing the robustness of the inner profile and the effectiveness of this simple cut in altering the two-halo signal. 
    Right: Comparison of fitted halo masses for the crowded subsamples (galaxies with a nearby BGS Bright galaxy) vs. non-crowded samples (galaxies without a nearby BGS Bright galaxy). 
    This split also shows $>1\sigma$ halo mass disagreement for only one stellar mass bin.
    }
    \label{fig:BGS-bright}
\end{figure*}

\subsection{The role of environment}
\label{sec:environment}
We further investigate the role of environment in two ways.
First, we compare the fitted halo masses from DESI-fiducial lensing signals computed with and without fiber completeness weights. As shown in Figure \ref{fig:BGS-bright} (left panel), the weights change the halo mass by significantly less than one-sigma for five of the six stellar mass bins, with the exception being the highest mass bin considered. However, four of the six bins have higher median halo masses when weights are included---this is in agreement with the results of \citet{Lange_2024} and with the expectation that targeting bias reduces the satellite fraction and thus the typical host halo mass of the spectroscopic sample.
We plan to revisit this issue with future DESI data, including weights for the full sample.  However, the change is not expected to be significant given current error bars, since the galaxies without weights comprise $<30\%$ of each stellar mass bin.

Second, we split the DESI-fiducial lenses based on whether or not they have at least one BGS Bright galaxy within 3 Mpc. We show the resulting signals for a combined dwarf bin and for the highest-mass bin in Figure \ref{fig:BGS-bright}. 
Qualitatively, this split has a similar effect to the first test.
Namely, while the signals in the one-halo regime (and halo masses) are largely consistent, every stellar mass bin shows a higher signal in the two-halo regime for the lenses with a nearby BGS Bright galaxy, demonstrating that this simple split is an effective proxy for environment. 
The inclusion of weights decreases the magnitude of this effect, since they effectively upweight the relatively crowded regions in the subsample without a BGS Bright galaxy within 3 Mpc.
The effect on the signal in the two-halo regime persists even when we use only a cut on angular separation (i.e., without any redshift information), highlighting the usefulness, even for photometric lenses, of splitting lenses on catalog-based projected distances to neighbors. 

Although the dependence of the lensing results on environment is clear, detailed interpretation will require more work. For example, spectroscopic success rate and fiber assignment effects prevent some relevant bright galaxies from inclusion in BGS Bright. We therefore leave 
full exploitation of environmental dependence and modeling of the satellites and centrals for future work, and focus here on the halo mass measurements.

\subsection{How do dwarf galaxy mass profiles vary with galaxy color, surface brightness, and size?}
\label{sec:DES_prop_splits}

\begin{figure*}
    \centering
    \includegraphics[width=0.9\textwidth]{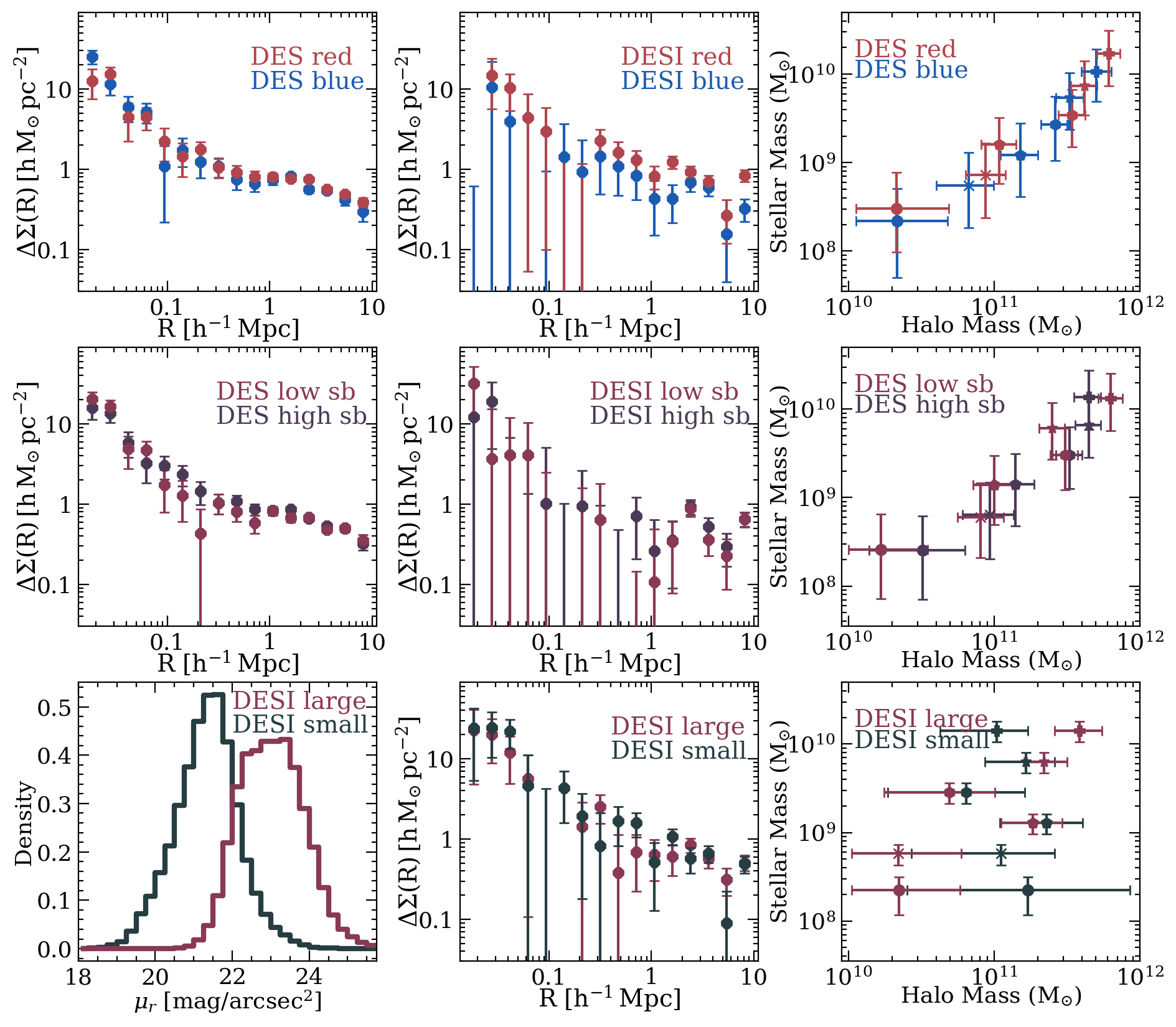}
    \caption{
    Upper panel:
    Weak lensing mass profiles for blue and red lens galaxies for the combined dwarf sample (\logmass~$<$9.25), computed using
    the photometric DES lens selection (left) and spectroscopic DESI-fiducial lens selection (middle).
    For this stellar mass range, the cutoff between blue and red is $g-r=0.38$.
    Using NFW fits to the inner regions ($R\lesssim0.15h^{-1}$Mpc) of the lensing profiles, we find consistent halo masses for the blue and red samples (right). 
    Middle panel: The same as above for low and high surface brightness subsamples, for a split at $\mu_r=23.24$. 
    Lower panel: DESI physical size split in the combined dwarf galaxy bin with stellar masses matched via weighting and lenses sampled such that the $g$--$r$ distributions are consistent. 
    While a physical size split is non-trivial for photometric lenses, surface brightness is an effective proxy (left).
    Halo mass is consistent within one-sigma for all but one stellar mass bin (right).
    }
    \label{fig:colorsb-splits}
\end{figure*}
Each stellar mass lens bin has sufficient signal-to-noise to test the dependence of the lensing profile on other photometric properties. We study the dependence on $g$--$r$ color, effective (angular) radius, and surface brightness. In this work, we consider each property split independently, but future higher-$S/N$ data will enable the isolation of more specific populations (e.g., galaxies that are compact, blue, and low-redshift). 

To divide each DES stellar mass bin by a given property, we sort the lenses by that property and split the sample in half. 
This sorting procedure often results in individual SOM cells contributing galaxies to both subsamples. Therefore, to compute a given subsample's stellar mass and redshift distributions, we only use the DESI calibrators that also satisfy the property cut.
This results in fewer than 20 calibrators for $<3\%$ of lenses. As a complement,
we also test the property dependence with DESI lenses. Although these signals are lower signal-to-noise, the comparison is cleaner given the individual stellar mass estimates. 
We perform the split on effective radius with DESI but not DES, since we need individual redshifts for converting from angular to physical radius.
For the other properties, rather than independently splitting the DESI samples in half, we use the cutoff value identified for the DES property splits. 

We present an example of a color split in Figure~\ref{fig:colorsb-splits} and the corresponding SHMRs for the blue and red subsamples of all six stellar mass bins. 
Every stellar mass bin shows consistency between the blue and red signals in the one-halo ($\lesssim 0.15 h^{-1} \rm Mpc$) regime and in the fitted halo mass. 
In the DES two-halo terms, the signals remain consistent within the uncertainties but the data points for the redder subsample tend to lie systematically above those of the bluer subsample.
The difference in the two-halo regime is clearer for the DESI galaxies (especially for higher-mass bins), which makes sense considering the wider photometric property distributions (Figure \ref{fig:exact_comparison}).
This higher two-halo signal is expected, as redder galaxies at a fixed stellar mass are more likely to be satellites of more massive host halos.
Furthermore, this discrepancy could explain the higher two-halo terms for DESI-fiducial signals compared to DES ones, since the DESI-fiducial sample tends to be redder (Figure \ref{fig:exact_comparison}).
As we show in the right panel of Figure~\ref{fig:colorsb-splits}, the red and blue galaxies in each stellar mass bin have consistent halo mass constraints.

We also see two-halo differences in some stellar mass bins when splitting by surface brightness and effective radius, but those splits correlate with color differences, making it unclear with DES lenses alone whether these inconsistencies are reducible to the effect of the color distribution. 
With DESI, we explore these property splits with more freedom to isolate differences in the property of interest, finding that only color has a clear effect in this scale and $S/N$ regime.

In Figure~\ref{fig:colorsb-splits} we show the lensing profiles for the low surface brightness and high surface brightness galaxies in the combined dwarf (\logmass~$<9.25$) bin.
After sampling the DESI lenses to match the $g$--$r$ distributions (in bins of width 0.1) and weighting them to get consistent stellar mass distributions (in bins of width 0.05 dex), the difference at larger scales disappears.
As we demonstrate in Appendix \ref{sec:stellar-mass-appendix}, this result is consistent across all four of the stellar mass approaches.

The subsamples from the surface brightness split are comparable to the physical size ones (Figure~\ref{fig:colorsb-splits}). With 
$g$--$r$ and stellar mass distributions matched, we see that there is insufficient evidence to conclude that galaxy effective radius correlates with the halo mass profile.
\section{Discussion}
\label{sec:discussion}
Our weak galaxy--galaxy lensing measurements open a path to constrain the low-mass regime of the stellar-to-halo mass relation and the full mass profiles of dwarf galaxies, where different dark matter models predict divergent behavior \citep{Wechsler2018}. 
Our analysis of profiles from stellar mass bins split by color, surface brightness, and size (Section \ref{sec:DES_prop_splits}) reveals no statistically significant dependence of the halo mass estimates on these galaxy properties. However, the two-halo term, which probes the satellite fraction, is sensitive to a range of factors.
Our comparison of the signals from DES and DESI lenses corroborates this conclusion and adds an important robustness test for our approach to select and calibrate photometric lenses. In particular, the signals for the photometric lenses are consistent at small scales with those from the spectroscopic lenses. Thus, at the level of precision of current measurements, sample incompleteness is not an obstacle to general conclusions about the SHMR. This finding is key considering the wealth of additional photometric survey data that we expect in the coming years.

In this section, we discuss our findings regarding the one-halo (Section \ref{sec:one-halo}) and two-halo (Section \ref{sec:two-halo}) ranges of the mass profiles. In Section \ref{sec:how-low}, we extend our dwarf galaxy selection to lower masses and smaller scales to present a simple forecast for an LSST analysis. 

\subsection{Robustness of the dwarf lensing halo masses}
\label{sec:one-halo}
By comparing photometric and spectroscopic lenses (Figure~\ref{fig:exact_comparison}) and splitting stellar mass bins by other galaxy properties (Figure~\ref{fig:colorsb-splits}), we find that the one-halo signal is robust, at least for the scales, signal-to-noise ratios, and parts of photometric space probed in this work. In particular, we find no statistically significant differences between any signals in a given stellar mass bin at scales of $\lesssim 0.15 h^{-1} \rm Mpc$. These inner points---and the resulting fitted halo masses---are also not sensitive to the specifics of the stellar mass distributions, as evidenced by the DES and DESI comparisons with similar mean stellar masses and notably different spreads (Figure~\ref{fig:exact_comparison}). 
This indicates that within our uncertainties, the one-halo regime is not sensitive to the satellite fraction, and is thus less susceptible to the various sources of incompleteness that affect the two-halo term. Specifically, the inner scales are not significantly affected by the incompleteness inherent to the spectroscopic sample, which is biased against dense regions and faint, red galaxies.
They are similarly robust to the incompleteness of the photometric sample, which is most prominent in the surface brightness and color distributions (Figure~\ref{fig:exact_comparison}). These results suggest that our halo mass constraints are robust.

We perform two additional tests to understand the effect of the complex selection of our DESI calibration sample.
First, we create versions of our SOM and DES signals using only DESI calibrators from BGS Bright, which is straightforwardly magnitude-limited ($r<19.5$), unlike the BGS Faint and LOWZ samples. Because of this magnitude limit, we impose a stricter magnitude cut throughout this test. Using these more restricted, cleaner samples, we find consistent signals as well as consistent fitted halo masses.

Likewise, we find that our analysis is not sensitive to redshift failures within DESI. 
This test is crucial since incompleteness from redshift failures could affect more than just satellite fraction---it could, for instance, introduce systematic biases in the calibrated stellar mass and redshift distributions.
Using a version of the SOM that includes the DESI galaxies that do not pass the \texttt{ZWARN} and \texttt{DELTACHI2} cuts (Section \ref{sec:desi-data}), we find that less than $5\%$ of calibrators in each stellar mass bin have failed redshifts (and therefore do not contribute to the calibration of the DES galaxies in that bin). We find that excluding SOM cells with $>10\%$ redshift failures does not significantly change the lensing mass profiles.

The insensitivity of the one-halo signal to selection effects in the context of this work offers insight into the galaxy--halo connection in a previously inaccessible, lower-mass regime.
Previous work by \cite{Desmond2017}, \cite{Hearin2019}, and \cite{Somerville2018} predicted that, at fixed stellar mass, smaller galaxies should live in more massive halos, although \cite{Hill_Mao2025} predict that the size dependence of halo properties is smaller for dwarf galaxies. Within our current uncertainties, we do not find a mass difference between smaller and larger galaxies. 
The relationship between dwarf galaxy color and $M_{\rm halo}/M_{*}$ remains an open question \citep[e.g.,][]{More2011,Rodriguez-Puebla2015,Zu2016,Moster2018,Wechsler2018,Behroozi2019}, and here we do not find a correlation.
Persistent halo mass consistency between red and blue galaxies in future, higher-precision measurements would indicate that the passive quenching model is disfavored \citep{Wechsler2018}.

It is entirely plausible that this one-halo insensitivity will break down at the smaller scales and reduced uncertainties accessible with LSST and other future lensing surveys. To fully assess this in future work, a full accounting of sample incompleteness to globally constrain halo masses and inner density profiles is needed. For now, we find that our calibration approach is effective and that our reported halo mass uncertainties likely encompass any scatter in halo mass at fixed stellar mass due to the galaxy--halo connection.

\subsection{Impact of selection effects on the two-halo profile}
\label{sec:two-halo}
The two-halo regime (scales $\gtrsim 0.15 h^{-1} \rm Mpc$) is clearly sensitive to spectroscopic targeting choices and photometric property incompleteness, although fully calibrated fiber completeness weights provide a promising avenue for mitigating the targeting effects. 
Because of this sensitivity, we do not attempt to provide physical conclusions about the fraction of our low-mass galaxies that are satellites.

Nonetheless, we speculate about the causes of relative differences between the two-halo signal strengths. 
We do not find sufficient evidence to conclude that surface brightness or physical size correlate with the strength of the signal in the two-halo regime (and therefore satellite fraction), although we expect such a difference may exist with higher $S/N$ measurements \citep[e.g., predictions of][]{Lehmann_2016}.
We consistently find that red galaxies have stronger two-halo terms at fixed stellar mass (Section \ref{sec:DES_prop_splits} and Figure~\ref{fig:colorsb-splits}), 
implying that red galaxies have a higher satellite fraction.
This finding is consistent with the expectation that satellite galaxies are more quenched than field galaxies, or alternatively with the expected relationship between quenched fraction and environmental density \citep[e.g.,][]{Peng2010,Wang2018}. 

This difference can also explain the slightly higher two-halo terms for DESI-fiducial signals, which include redder lenses than the photometric DES samples (Figure \ref{fig:exact_comparison}). The decreased two-halo strength of the DESI-SOM signals supports this interpretation, since these samples are based on lenses with stellar mass and photometric property distributions matched to DES. We expect that the inclusion of weights for the full sample will yield fully consistent DES and DESI-SOM signals and will explore that in future work.

\subsection{Dwarf lensing with next-generation imaging}
\label{sec:how-low}
\begin{figure}
    \centering
    \includegraphics[width=1\columnwidth]{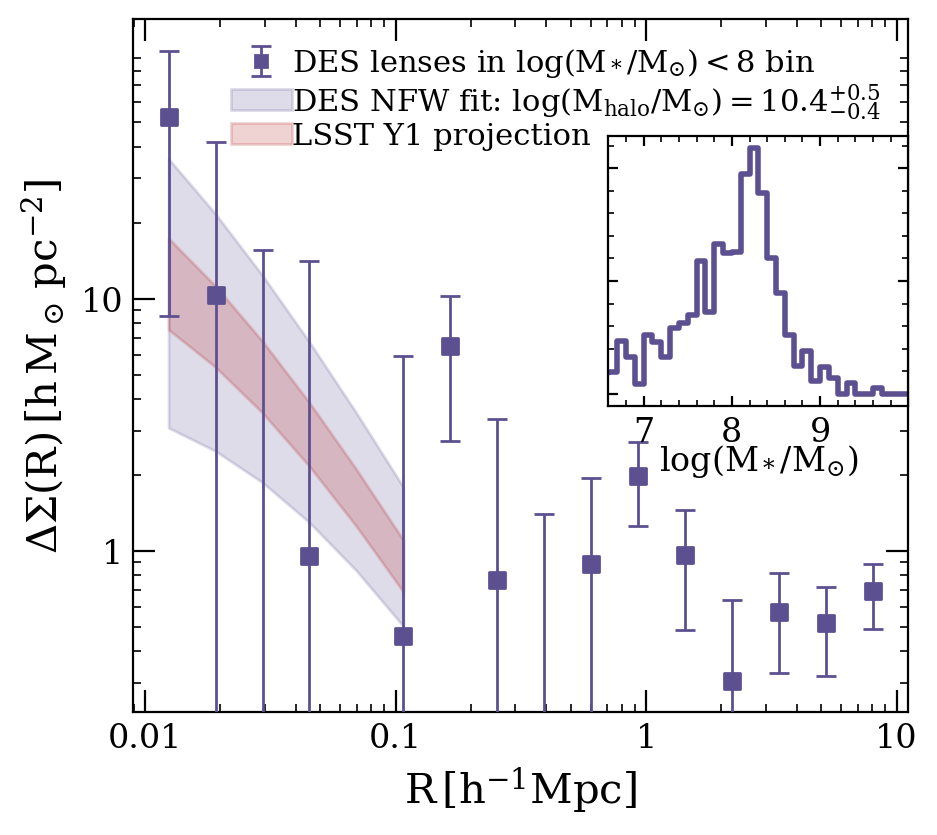}
    \caption{Lensing signal for DES lenses in SOM cells with $\rm \langle$\logmass$\rangle<8$ (purple), with the radial binning extending to 10 $h^{-1}$ kpc. The gray shading shows the one-sigma uncertainty on the NFW fit from these measurements. 
    The stellar mass distribution is inset.
    Although this bin still has signal, there is no added constraining power from the NFW fit (Table~\ref{tab:modeling}).
    The red shading represents the forecasted maximum LSST Y1 uncertainty for this mass bin. We account for the increased area and source number density but LSST affords additional improvements that will further decrease this uncertainty (Section~\ref{sec:how-low}). 
    }
    \label{fig:low}
\end{figure}
In this section, we push our sample to the lowest possible masses and estimate what will be possible for future data.
In Figure~\ref{fig:low}, we plot the \logmass$<10^{8} M_{\odot}$ signal, extending the radial binning to 10 $h^{-1}$kpc, and the corresponding NFW fit. This fit has a higher halo mass median than that of the $<10^{8.55} M_{\odot}$ bin
but larger uncertainties, adding no additional constraining power. 
We include the stellar mass and halo mass medians and one-sigma uncertainties of this sample in Table~\ref{tab:modeling}.

Our measurements mark an important first step, and the precision will improve dramatically with future surveys. Here we illustrate the forecast for LSST Year 1, while highlighting that Euclid and Roman will likewise provide transformative gains in dwarf galaxy lensing. In Figure~\ref{fig:low} we show forecasts for the dramatically smaller measurement uncertainty expected within the first year of LSST. To construct these, we simply scale down the measurement variance using the ratios between the current DES (5.6 galaxies/arcmin$^2$) and expected LSST Y1 (10 galaxies/acrmin$^2$) source densities and between the on-sky area of DES (5,000~degrees$^2$) and LSST (18,000~degrees$^2$), following \citet{LSST_SRD}. We then also account for the different typical source redshifts using Equation \ref{eq:uncorrected}, since the square of the distance-based prefactor is proportional to the variance. We note that this forecast underestimates the power of dwarf lensing with LSST in three critical ways: \\
(1) We do not account for the fact that deeper LSST data will result in a higher density of galaxies. Here we simply assume the DES dwarf galaxy density, rescaled to a larger area. \\
(2) The higher-density background sample achievable with LSST will enable galaxy--galaxy lensing measurements on smaller radial scales, of key science interest for studies of DM physics including DM self-interactions. \\
(3) LSST data includes wider wavelength coverage ($ugrizy$) than the DES data used here ($griz$). The six-band information has the potential to address one of the main limitations of this work: it can potentially reduce the scatter within each SOM cell and more precisely characterize the stellar mass of individual galaxies. This information could also help extend this study to go beyond stellar mass and redshift to include additional physical properties like the star formation rate in the SOM decomposition.

Paired with the smaller scales that will become accessible thanks to the depth, area, and resolution of LSST, these uncertainties are sufficiently small to allow for differentiation between core and cusp profiles as well as detailed studies of the dependence of the low-mass stellar mass--halo mass relation on galaxy properties.

\section{Conclusions}
\label{sec:conclusion}
We use an unprecedented sample of DESI spectroscopy to study the halo mass profiles of low-mass galaxies. Our sample is split into six stellar mass bins with median masses ranging from \logmass~$\sim8.3$--$10.1$, thus providing a new view into the halo profiles of dwarf galaxies.
We use two techniques and lens samples. The first uses DESI spectroscopic galaxies as lenses.
The second uses a subset of those galaxies for a calibration sample to build a substantially larger photometric dwarf galaxy sample of 700,000 galaxies from DES imaging, using an improved version of the methodology presented in \citetalias{Thornton2024}.
We constrain the halo mass of each subsample by fitting the small scales of the lensing profiles ($\lesssim 0.15 h^{-1} $Mpc). 
Our key findings are as follows: 
\begin{enumerate}
\item We have presented the most precise spectroscopic dwarf galaxy weak lensing measurements to date.
We detect the excess surface mass density profile of spectroscopically identified dwarf galaxies in DESI with combined $S/N\sim14$. Using six stellar mass bins, three in the dwarf galaxy regime and three at slightly higher masses, we constrain the halo masses and the SHMR from $\log_{10} M_{\rm {halo}}/M_\odot\sim10.7-11.6$. These constraints are robust to stellar mass estimation, fiber completeness, and weak lensing systematics. 
\item We are able to select and characterize a substantially larger sample of dwarf galaxies from DES imaging using an improved methodology compared to \citetalias{Thornton2024} and a 
$\times$$\sim$100 larger calibration sample from DESI.  These produce high-significance mass profile measurements with combined $S/N\sim38$ in the dwarf galaxy regime, and allow us to constrain the SHMR over $\log M_{\rm h}/M_\odot\sim10.2-11.7$. Through a detailed comparison of the measurements using the two approaches, our dual approach demonstrates that one-halo mass measurements are robust against systematic and sample selection effects, providing secure constraints on dwarf galaxy halo masses across both spectroscopic and photometric samples.
\item The larger photometric sample opens the possibility of even lower mass detections, with a $S/N\gtrsim5$ for an additional $\log_{10} M_*/M_\odot < 8$ bin, representing the lowest-mass lensing measurement to date, in combination with the detections achieved for the bins defined as \logmass~$< 8.55$, 8.55$-$8.9, and 8.9$-$9.25, with $S/N$ of 11, 20, and 30, respectively. 
\item The halo mass constraints, determined from the one-halo term of the measurements, are remarkably robust to different galaxy properties. We study how the measurements and halo mass estimates depend on galaxy color, size, and surface brightness at fixed stellar mass and find no significant trends.
\item We find that the larger-scale, two-halo measurements are sensitive to galaxy properties such as color and environment. This points to exciting opportunities for future higher-$S/N$ data to probe dwarf satellite fractions, once selection effects including color, environment, and target incompleteness are well controlled.
\end{enumerate}
This work presents the most precise dwarf lensing measurements to date and extends into a new low-mass frontier for galaxy--galaxy lensing. Our results demonstrate that DESI spectroscopy and DES-calibrated photometry already deliver high-significance dwarf lensing mass profiles, establishing a robust methodology that next-generation spectroscopic and lensing surveys will amplify by orders of magnitude.

Galaxy--galaxy lensing measurements of dwarf galaxies have the potential to also constrain the full galaxy--halo connection for a given sample, including the satellite fraction and scatter in the stellar mass--halo mass relation \citep[see the discussion in][for example]{Thornton2024}. To achieve this requires further work, including a clear understanding the selection effects in the dwarf sample. In this context it is important to note that our photometric dwarf methodology preferentially selects a color-biased sample. We demonstrate here that this selection does not significantly impact the measurements in the one-halo regime, but our results for different galaxy color and environment selections at a given stellar mass show that these properties impact the two-halo measurements. Future analyses will require modeling that can account for these completeness effects. 
A more sophisticated simulation-based modeling of the halo mass profiles, building upon \citet{Thornton2024} with a more comprehensive galaxy-property dependence beyond stellar mass, will help separate effects of dark matter and of halo occupation from those of galaxy formation.
To fully realize the statistical power of future surveys like LSST, higher-resolution, larger-volume simulations than are currently available will be required; ideally such simulations should match or exceed the volume probed by upcoming data while having sufficient resolution to fully track halo substructure, and should also explore the impact of baryonic and dark matter physics.

Related work by To et al (2025) has measured the lensing profile of a similar DESI spectroscopic sample with an independent lensing analysis based on the DECADE survey. Their work uses the DESI BGS Bright dwarf galaxy sample, and fully characterizes the impact of the individual inverse probability weights on the results. They also perform a full simulation-based analysis to constrain the galaxy--halo connection. We plan a full comparison of these independent studies in future work. 

This work provides a roadmap for dwarf lensing with expanded spectroscopic coverage from DESI, the Prime Focus Spectrograph \citep[][]{Takada2014}, and 4MOST \citep{deJong2019}, and deeper lensing data from  LSST \citep[][]{Ivezic2019},
Euclid \citep{Laureijs2011}, and Roman \citep{Roman2025}. The methodology for selecting dwarf lenses from photometric surveys is scalable for next-generation imaging, will give more precise stellar mass distributions with additional photometric bands, and will reveal lower mass halos with deeper data. Future weak lensing data will amplify our approach and provide the unique power to test the nature of dark matter, constrain galaxy evolution models, and determine the galaxy--halo connection for the lowest-mass galaxies.

\begin{acknowledgements}
HT acknowledges support by the National Science Foundation Graduate Research Fellowship Program under Grant DGE-2039656. 
Any opinions, findings, and conclusions or recommendations expressed in this material are those of the authors and do not necessarily reflect the views of the National Science Foundation.

This work received support from the Kavli Institute for Particle Astrophysics and Cosmology at Stanford University and SLAC National Accelerator Laboratory and from the U.S. Department of Energy under contract number DE-AC02-76SF00515 to SLAC National Accelerator Laboratory.

\textbf{DESI: }This material is based upon work supported by the U.S. Department of Energy (DOE), Office of Science, Office of High-Energy Physics, under Contract No. DE–AC02–05CH11231, and by the National Energy Research Scientific Computing Center, a DOE Office of Science User Facility under the same contract. Additional support for DESI was provided by the U.S. National Science Foundation (NSF), Division of Astronomical Sciences under Contract No. AST-0950945 to the NSF’s National Optical-Infrared Astronomy Research Laboratory; the Science and Technology Facilities Council of the United Kingdom; the Gordon and Betty Moore Foundation; the Heising-Simons Foundation; the French Alternative Energies and Atomic Energy Commission (CEA); the National Council of Humanities, Science and Technology of Mexico (CONAHCYT); the Ministry of Science, Innovation and Universities of Spain (MICIU/AEI/10.13039/501100011033), and by the DESI Member Institutions: \url{https://www.desi.lbl.gov/collaborating-institutions}. Any opinions, findings, and conclusions or recommendations expressed in this material are those of the author(s) and do not necessarily reflect the views of the U. S. National Science Foundation, the U. S. Department of Energy, or any of the listed funding agencies.

The authors are honored to be permitted to conduct scientific research on I'oligam Du'ag (Kitt Peak), a mountain with particular significance to the Tohono O’odham Nation.

\noindent\textbf{Imaging surveys: }

\textit{KiDS-1000: }Based on observations made with ESO Telescopes at the La Silla Paranal Observatory under programme IDs 177.A-3016, 177.A-3017, 177.A-3018 and 179.A-2004, and on data products produced by the KiDS consortium. The KiDS production team acknowledges support from: Deutsche Forschungsgemeinschaft, ERC, NOVA and NWO-M grants; Target; the University of Padova, and the University Federico II (Naples).

\textit{DES Y3: }This project used public archival data from the Dark Energy Survey (DES). Funding for the DES Projects has been provided by the U.S. Department of Energy, the U.S. National Science Foundation, the Ministry of Science and Education of Spain, the Science and Technology FacilitiesCouncil of the United Kingdom, the Higher Education Funding Council for England, the National Center for Supercomputing Applications at the University of Illinois at Urbana-Champaign, the Kavli Institute of Cosmological Physics at the University of Chicago, the Center for Cosmology and Astro-Particle Physics at the Ohio State University, the Mitchell Institute for Fundamental Physics and Astronomy at Texas A\&M University, Financiadora de Estudos e Projetos, Funda{\c c}{\~a}o Carlos Chagas Filho de Amparo {\`a} Pesquisa do Estado do Rio de Janeiro, Conselho Nacional de Desenvolvimento Cient{\'i}fico e Tecnol{\'o}gico and the Minist{\'e}rio da Ci{\^e}ncia, Tecnologia e Inova{\c c}{\~a}o, the Deutsche Forschungsgemeinschaft, and the Collaborating Institutions in the Dark Energy Survey.
The Collaborating Institutions are Argonne National Laboratory, the University of California at Santa Cruz, the University of Cambridge, Centro de Investigaciones Energ{\'e}ticas, Medioambientales y Tecnol{\'o}gicas-Madrid, the University of Chicago, University College London, the DES-Brazil Consortium, the University of Edinburgh, the Eidgen{\"o}ssische Technische Hochschule (ETH) Z{\"u}rich,  Fermi National Accelerator Laboratory, the University of Illinois at Urbana-Champaign, the Institut de Ci{\`e}ncies de l'Espai (IEEC/CSIC), the Institut de F{\'i}sica d'Altes Energies, Lawrence Berkeley National Laboratory, the Ludwig-Maximilians Universit{\"a}t M{\"u}nchen and the associated Excellence Cluster Universe, the University of Michigan, the National Optical Astronomy Observatory, the University of Nottingham, The Ohio State University, the OzDES Membership Consortium, the University of Pennsylvania, the University of Portsmouth, SLAC National Accelerator Laboratory, Stanford University, the University of Sussex, and Texas A\&M University.
Based in part on observations at Cerro Tololo Inter-American Observatory, National Optical Astronomy Observatory, which is operated by the Association of Universities for Research in Astronomy (AURA) under a cooperative agreement with the National Science Foundation.

\textit{SDSS: }Funding for the SDSS and SDSS-II has been provided by the Alfred P. Sloan Foundation, the Participating Institutions, the National Science Foundation, the U.S. Department of Energy, the National Aeronautics and Space Administration, the Japanese Monbukagakusho, the Max Planck Society, and the Higher Education Funding Council for England. The SDSS Web Site is http://www.sdss.org/.

The SDSS is managed by the Astrophysical Research Consortium for the Participating Institutions. The Participating Institutions are the American Museum of Natural History, Astrophysical Institute Potsdam, University of Basel, University of Cambridge, Case Western Reserve University, University of Chicago, Drexel University, Fermilab, the Institute for Advanced Study, the Japan Participation Group, Johns Hopkins University, the Joint Institute for Nuclear Astrophysics, the Kavli Institute for Particle Astrophysics and Cosmology, the Korean Scientist Group, the Chinese Academy of Sciences (LAMOST), Los Alamos National Laboratory, the Max-Planck-Institute for Astronomy (MPIA), the Max-Planck-Institute for Astrophysics (MPA), New Mexico State University, Ohio State University, University of Pittsburgh, University of Portsmouth, Princeton University, the United States Naval Observatory, and the University of Washington.

\end{acknowledgements}

\newpage
\appendix
\section{Stellar Masses}
\begin{figure*}
    \centering
    \includegraphics[width=1\linewidth]{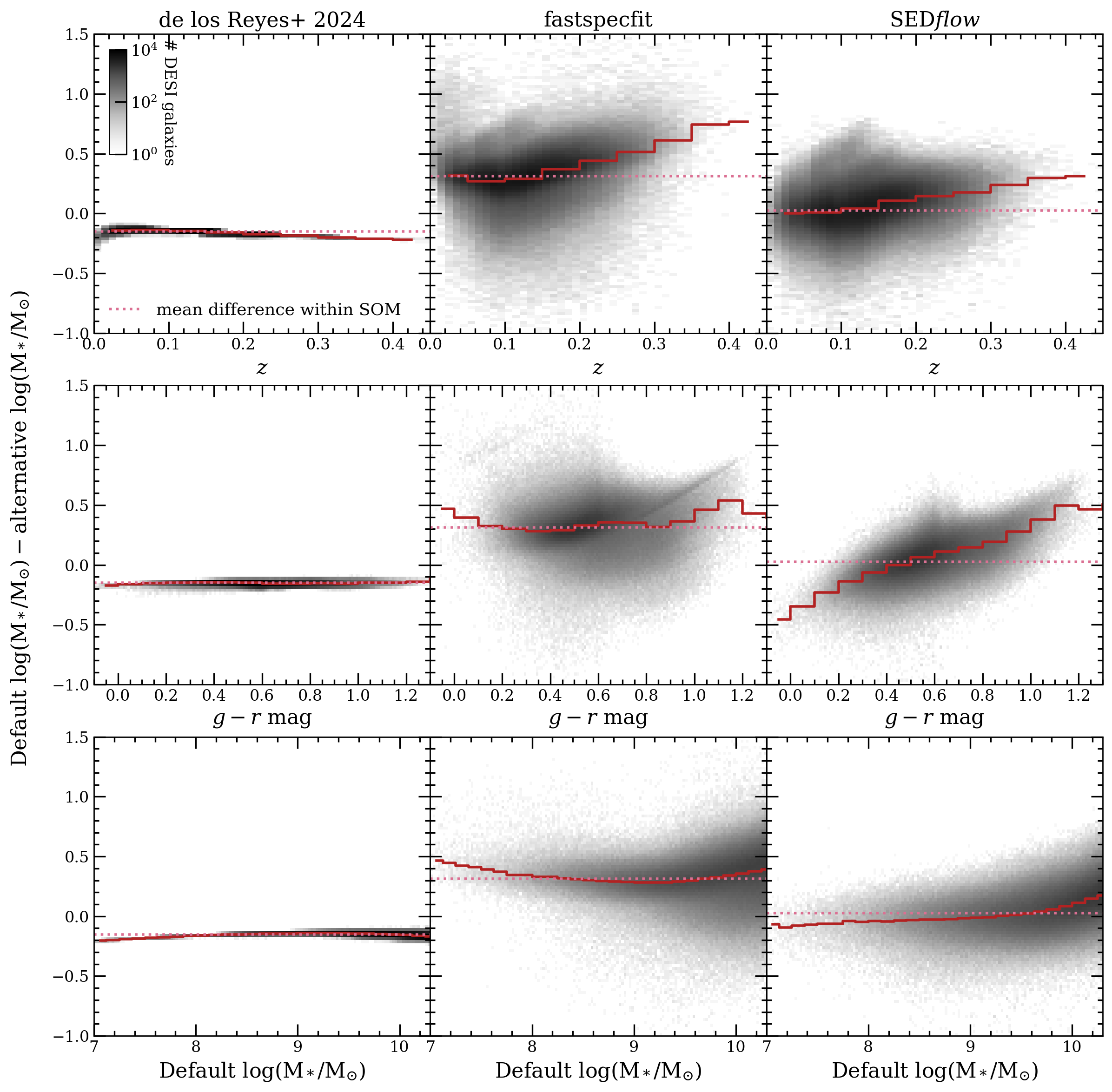}
    \caption{Comparison between the default stellar mass estimate used here and three others available within DESI that are expected to work reasonably well for low-mass and low-redshift galaxies. Each column corresponds to one of the three approaches and each row shows the difference between the default method and the alternative estimate as a function of redshift (top), color (middle), and the default mass (bottom). The red step functions track the median values while the horizontal pink dotted lines mark the overall mean difference within the SOM. This value tends to sit closer to zero since the photometric cuts remove the red, high-redshift, and high-mass galaxies that are most likely to have significant disagreement between the stellar mass estimates.
    }
    \label{fig:stellar-masses}
\end{figure*}
\label{sec:stellar-mass-appendix}

As we move to more precise lensing measurements and incorporate investigations into the role of properties beyond mass, it is important to consider the complexity of stellar mass estimates. 
We investigate four sets of stellar mass estimates for the DESI sample.
We use these approaches to check the robustness of our results, especially in the case of investigations into how other galaxy properties affect the lensing profile.
\begin{enumerate}
    \item The first set of estimates uses the SAGA approach, designed for low-mass, low-redshift galaxies. As outlined in \cite{Mao2021}, the relation uses \cite{Chilingarian2010} $k$-corrections with \cite{Bell2003} mass-to-light ratios adapted for a Kroupa IMF \citep{Kroupa2001}: \logmass~$ = 1.254 + 1.98(g-r) - 0.4M_r$. Since this relation is designed for SDSS magnitudes, we first use the \cite{Sevilla-Noarbe2021} conversions to go from DES to SDSS $g$ and $r$ magnitudes. These conversions only affect the mass estimates by $\sim$0.025 dex.
    \item The relation presented in \cite{delosReyes2024} is similar but explicitly tailored to be accurate for the masses and star formation histories of dwarf galaxies. The fit uses mock observations applied to both semi-empirical models and cosmological baryonic zoom-in simulations. The residuals of masses estimated from these models tend to sit within 0.2 dex of \logmass~$ = 1.433(g-r) + 0.00153M_g^2 - 0.335M_g + 2.072$.
    However, the fit only incorporates galaxies with \logmass~$<10$ and \cite{delosReyes2024} note that errors are expected for galaxies with \logmass~$>9$. Neither of these two relations can handle galaxies at $z>0.5$ (of which there are 5,317 in the DESI calibration sample) because the $k$-corrections are no longer effective at that point.
    \item In contrast, \texttt{fastspecfit} \citep{Moustakas2023} estimates stellar masses using all the photometry and spectroscopy available through DESI. 
    Of course, this approach still requires assumptions, including about the IMF (\citealt{Chabrier2003} with maximum mass 100 $\rm M_{\odot}$), metallicity (0.1 to 2 times solar), dust (power-law with slope of $-$0.7 and \citealt{Draine2007} emission spectra), and the lack of a significant AGN component.
    In this paper, we use \texttt{fastspecfit} v2.1.
    \item Intermediate to these approaches is SED$flow$ \citep{Hahn2022}, which uses PROVABGS \citep{Prova2023} SED models to train a neural network that can then quickly estimate the posterior. As a result, like the simpler relations, SED$flow$ will remain a feasible mass estimate approach for upcoming surveys.
    Both \texttt{fastspecfit} and SED$flow$ use non-parametric star formation histories and are designed to still work well for low-redshift/low-mass galaxies.
    \cite{Hahn2022} note that SED$flow$ successfully recovers stellar masses and any inaccuracy does not correlate with colors or magnitudes, but we note that \cite{delosReyes2024} found offsets in the dwarf galaxy regime using this method.
\end{enumerate}
The two relations using redshift and two magnitudes differ by $\sim$0.15 dex, while \texttt{fastspecfit} estimates are offset, on average, by 0.3 dex from the relation used in SAGA, calling into question whether it is sufficient to quote 0.2 dex as the systematic uncertainty. 
The uncertainty is comparably high according to SED$flow$ but the average difference is zero in the context of the SOM.

We investigate the differences between four stellar mass estimate approaches to provide more accurate error bars in our stellar vs. halo mass results and to understand whether the approaches all qualitatively agree in the context of the property splits (Section \ref{sec:DES_prop_splits}). Here, we expand on this comparison and the resulting justification for the 0.3 dex systematic uncertainty added in Figures \ref{fig:SHMR} and \ref{fig:pilot-SHMR}.

In Figure~\ref{fig:stellar-masses}, we plot the differences between our default method (used in \citetalias{Thornton2024} and \citealt{Mao2021,Mao2024}) and the other three as a function of redshift, $g$--$r$ color, and the default stellar mass. These three variables are clearly correlated with these approaches' disagreements, while other photometric properties (like surface brightness) are not. These plots include the entire DESI-fiducial lens sample (i.e., DESI Data Release 1 galaxies with $z<0.5$ and \logmass~$<10.3$), although the \texttt{fastspecfit} and SED$flow$ estimates are not available for some of the included galaxies.
The red step functions trace the median differences in the default and alternative stellar masses. These medians are nearly constant (at $-0.15$ dex) for the \cite{delosReyes2024} approach, which, like the default one, uses only redshift and $g$- and $r$-band magnitudes. The other two methods tend to favor lower stellar mass estimates, with the disagreement increasing at higher redshifts, higher stellar masses, and for redder galaxies; these three variables are, of course, correlated with each other. The scatter is significant, though, so it is not straightforward to predict the exact uncertainty using the properties explored in this work.

In the context of the SOM, however, the $\gtrsim0.5$ dex disagreements are significantly less relevant, since the photometric cuts remove many of the high-mass, high-redshift, red galaxies. We highlight this discrepancy by also plotting the overall mean disagreement within the SOM; \texttt{fastspecfit} then dominates the resulting uncertainty, with a mean of 0.3 dex. When presenting stellar vs. halo mass constraints, we adopt 0.3 dex as an additional term added in quadrature to the one-sigma error bar from the default approach's distribution of stellar masses.
We note for future work that this value can be finer-grained; one might use the individual objects' uncertainties, mean uncertainties from SOM cells, or the mean uncertainty for a given stellar mass bin. We find in this context, though, that the six stellar mass bins all have $ \langle \mathrm{default} \, \log_{10}(M_*/M_{\odot})-\texttt{fastspecfit} \, \log_{10}(M_*/M_{\odot}) \rangle$ values that round to 0.3 dex. 
Moreover, this added uncertainty is approximate here since all four estimates share several assumptions that may not hold for our full sample, such as the metallicity range and abundance ratios.

\begin{figure}
    \centering
    \includegraphics[width=1\linewidth]{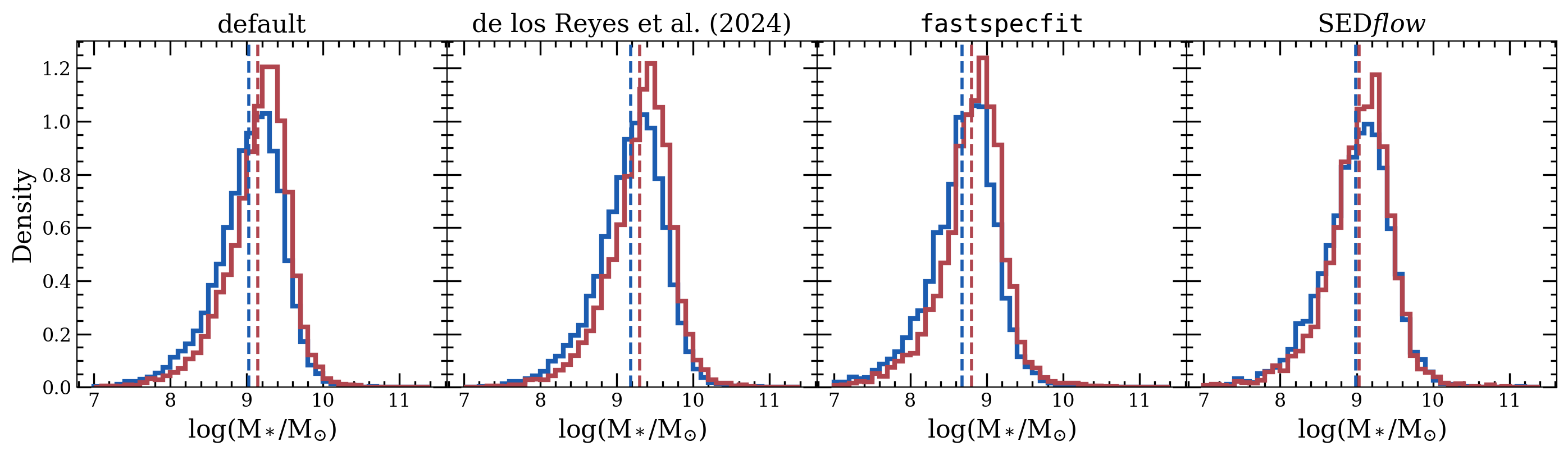}
    \caption{Stellar mass distributions for the blue and red subsamples of the $8.9-9.25$ bin according to the four mass approaches. 
    Vertical dashed lines mark the mean stellar masses.
    Although the differences between the mass approaches' estimates are often correlated with color, they qualitatively agree that the color splits yield comparable mass distributions with a slightly higher mean stellar mass for the red galaxies.}
    \label{fig:gr-masses}
\end{figure}

Because of the correlation with color, it is especially important to check whether the red and blue stellar mass distributions are similar relative to each other, regardless of method. We show an example of this comparison in Figure~\ref{fig:gr-masses}. The red distribution's mean is always comparable or slightly higher than that of the blue distribution, meaning the interpretation is consistent across the methods, at least for the purpose of this work.

\begin{figure*}
    \centering
    \includegraphics[width=1\linewidth]{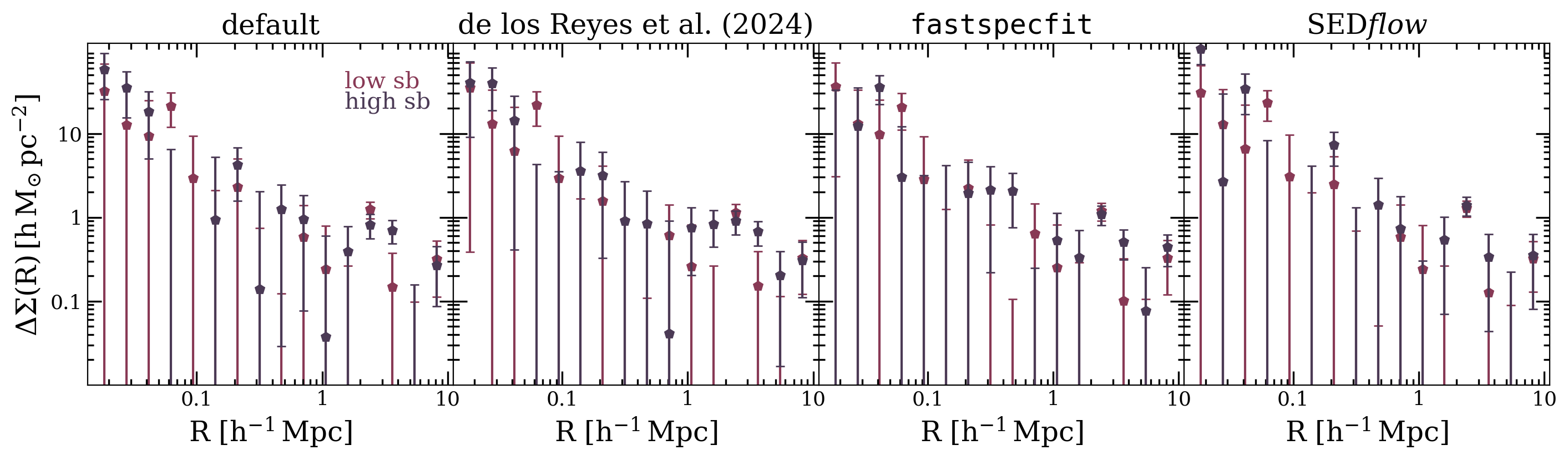}
    \caption{
    DESI surface brightness split in the $8.9<$~\logmass~$<9.25$ bin with stellar masses matched. 
    For each panel, we sample from the two surface brightness bins to ensure matched $g$--$r$ distributions.
    The four panels each use a different stellar mass approach for the mass weighting. Regardless of mass estimate approach, we see that the high and low surface brightness galaxies have consistent profiles once their color and stellar mass distributions are matched.}
    \label{fig:sb-split}
\end{figure*}

Similarly, we repeat with all four methods the stellar mass distribution matching that we do for DESI profiles that we split on galaxy properties. In Figure~\ref{fig:sb-split}, we plot a representative example demonstrating that the mass-matched lensing signals are consistent regardless of method.
All in all, while we find that the stellar mass uncertainty is likely higher than the 0.2 dex reported for the default and \cite{delosReyes2024} approaches, our results are otherwise unaffected by the estimate approaches' disagreements. Advancements in constraints on the SHMR will ideally leverage smaller error bars achieved not just through improving the SOM method but also through better understanding these stellar mass uncertainties and how they change throughout the photometric space of interest.

\section{Direct Comparison to T24}
\label{sec:pilot-comparison-appendix}
\begin{figure*}
    \centering
    \includegraphics[width=1\linewidth]{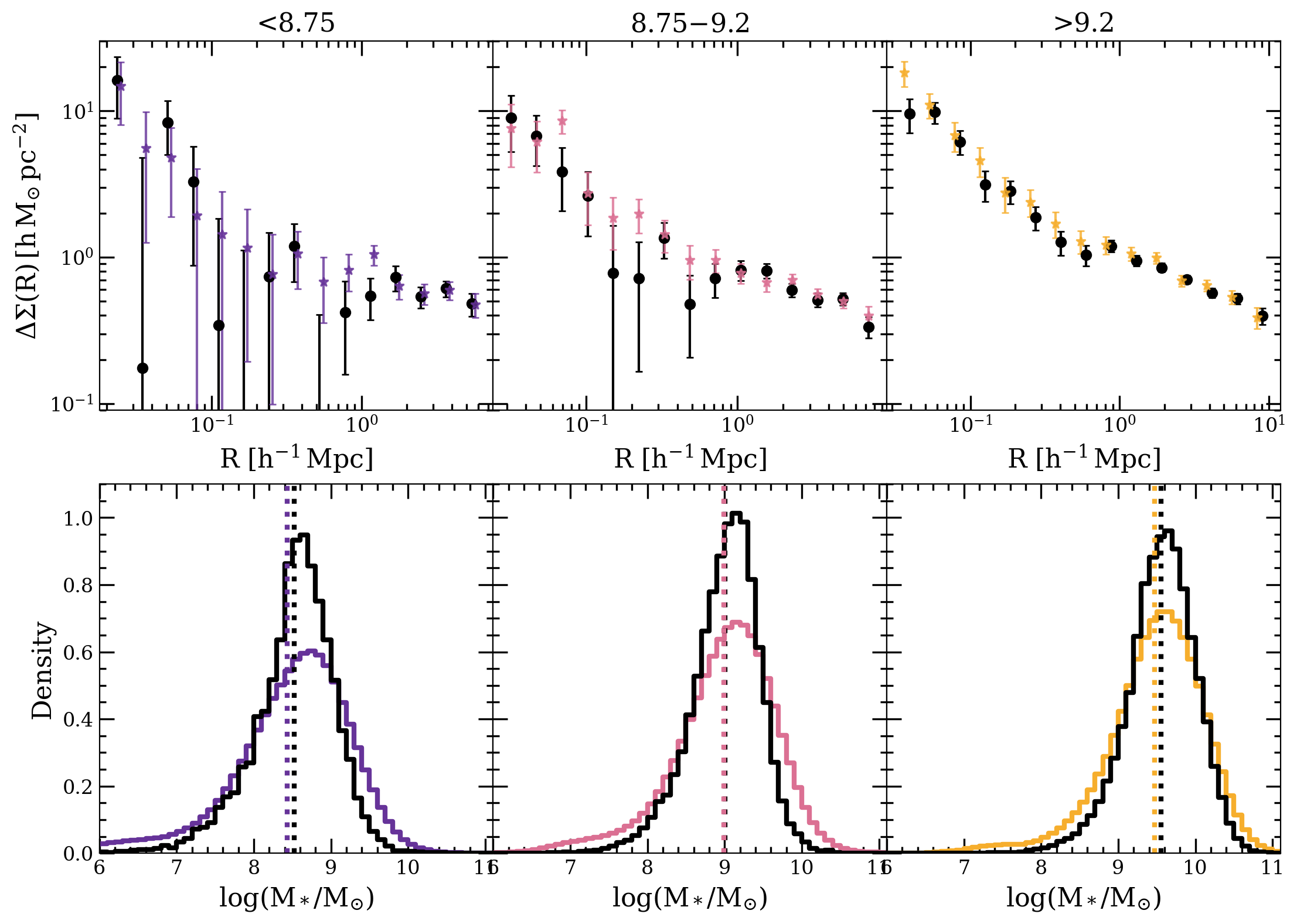}
    \caption{
    Comparison between the lensing signals (top) and stellar mass distributions (bottom) from \citetalias{Thornton2024} and this work using the updated SOM and DESI calibrators but the same photometric cuts as the T24 pilot study. The updated versions are in black, indicating narrower distributions and slightly higher masses for the same SOM bins. Vertical dotted lines mark the mean masses.}
    \label{fig:pilot-comparison}
\end{figure*}

Here we expand on our comparison to \citetalias{Thornton2024} by using their stellar mass bounds ($<8.75$, $8.75-9.2$, $9.2-11$) and radial binning approach. We investigate two sets of comparisons,
one using our updated SOM and one using the \citetalias{Thornton2024} photometric cuts, such that we can expect consistency even in the two-halo term. In Figure~\ref{fig:pilot-comparison}, we compare the latter (i.e., the $z<0.01$-complete signals) to those presented in \citetalias{Thornton2024}.
While the calibrated mean stellar masses are comparable, our updated SOM yields a narrower stellar mass distribution in each bin. The updated lens signals have noticeably larger error bars, especially in the two lower-mass bins, partially because of the stricter quality cuts that removed more of the DES objects.

\begin{figure}
    \centering
    \includegraphics[width=0.5\linewidth]{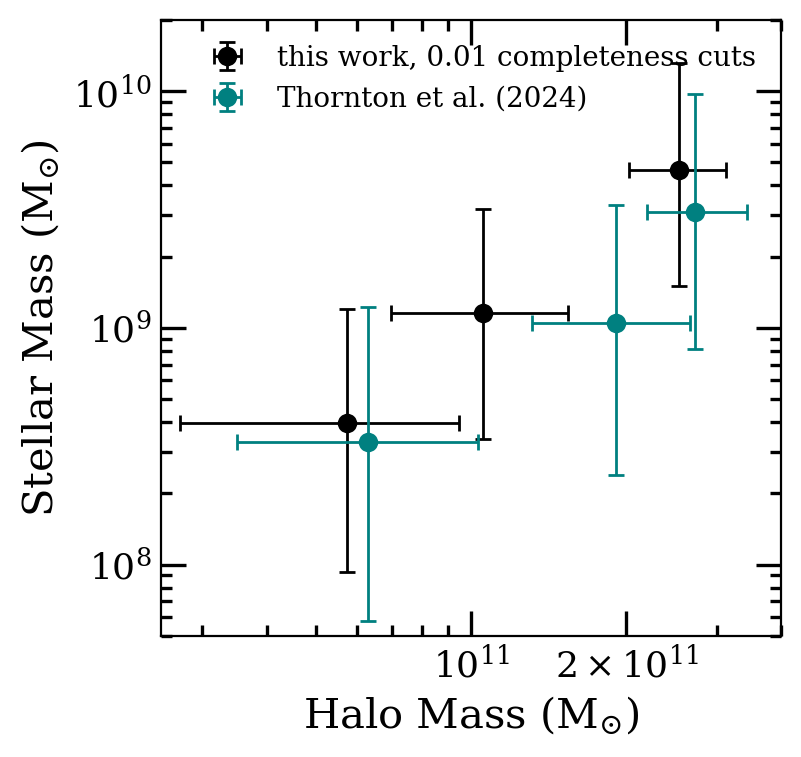}
    \caption{Stellar vs. NFW halo mass fits for the three stellar mass bins from \citetalias{Thornton2024} (teal) and for the analogously defined bins with the same photometric cuts but the updated SOM. In all three cases, the halo masses are consistent but slightly lower in this work. 
    }
    \label{fig:pilot-SHMR}
\end{figure}

To quantify the comparison, we perform NFW fits (Figure~\ref{fig:pilot-SHMR}), finding consistent halo masses but a systematic offset, which comes from DESI tending to calibrate SOM cells to higher mean redshifts and stellar masses. This effect is clearest when we compare the SAGA and DESI calibrations for the \citetalias{Thornton2024} SOM. The mean disagreement between DESI and SAGA is highest at low stellar mass and low redshift. 
The offset is also apparent in Figure~\ref{fig:SHMR}, demonstrating that the spectroscopic lensing also finds this slight disagreement.

\bibliography{paper.bib}
\bibliographystyle{aasjournal}

\end{document}